\documentclass[sigconf]{acmart}

\AtBeginDocument{%
  \providecommand\BibTeX{{%
    Bib\TeX}}}

\copyrightyear{2023} 
\acmYear{2023} 
\setcopyright{acmlicensed}\acmConference[SC '23]{The International Conference for High Performance Computing, Networking, Storage and Analysis}{November 12--17, 2023}{Denver, CO, USA}
\acmBooktitle{The International Conference for High Performance Computing, Networking, Storage and Analysis (SC '23), November 12--17, 2023, Denver, CO, USA}
\acmPrice{15.00}
\acmDOI{10.1145/3581784.3607034}
\acmISBN{979-8-4007-0109-2/23/11}

\usepackage{amsmath,amsfonts}
\usepackage[ruled,vlined]{algorithm2e}
\usepackage{tabularx}
\usepackage{subfig}
\usepackage{graphicx}
\usepackage{physics}
\usepackage{textcomp}
\usepackage{fancyhdr}
\usepackage{enumitem}
\usepackage{afterpage}
\usepackage{array}
\usepackage{bm}
\usepackage{float}
\usepackage[normalem]{ulem}
\usepackage{caption}
\usepackage{environ}
\usepackage{makecell}
\usepackage{textcomp}
\usepackage{booktabs}
\usepackage{lipsum}
\usepackage{pifont}
\newcolumntype{P}[1]{>{\centering\arraybackslash}p{#1}}
\colorlet{shadecolor}{gray!20}
\def\BibTeX{{\rm B\kern-.05em{\sc i\kern-.025em b}\kern-.08em
    T\kern-.1667em\lower.7ex\hbox{E}\kern-.125emX}}

\newcommand{\sol}{\textsc{Clover}}
\newcommand{\blover}{\textsc{Blover}}
\newcommand{\oracle}{\textsc{ORACLE}}
\newcommand{\base}{\textsc{BASE}}
\newcommand{\opt}{CO\textsubscript{2}OPT}
\newcommand{\coo}{CO\textsubscript{2}}

\usepackage[format=plain,labelfont=it,labelfont=bf,labelsep=period,textfont=it]{caption}

\usepackage{tcolorbox,tikz}
\tcbuselibrary{skins}
\tcbuselibrary{breakable}

\newtcolorbox{mybox}[3][]
{
  breakable, 
  enhanced,
  colback=white!95!black,
  colframe=black,
  boxsep=-0.5mm,
  sharp corners,
  boxrule=0.5pt,
  left=2mm,
  #1,
}

\renewcommand\footnotetextcopyrightpermission[1]{}

\newtheorem{definition}{Definition}

\begin{document}

\title{\sol{}: Toward Sustainable AI with Carbon-Aware Machine Learning Inference Service}

\author{Baolin Li}
\affiliation{%
  \institution{Northeastern University}
  \country{}}

\author{Siddharth Samsi}
\affiliation{%
  \institution{MIT Lincoln Laboratory}
  \country{}}

\author{Vijay Gadepally}
\affiliation{%
  \institution{MIT Lincoln Laboratory}
  \country{}}

\author{Devesh Tiwari}
\affiliation{%
  \institution{Northeastern University}
  \country{}}

\begin{abstract}
This paper presents a solution to the challenge of mitigating carbon emissions from hosting large-scale machine learning (ML) inference services. ML inference is critical to modern technology products, but it is also a significant contributor to carbon footprint. We introduce, \sol{}~\footnote{\sol{} has been accepted at the 2023 ACM/IEEE International Conference for High Performance
Computing, Networking, Storage, and Analysis (SC '23)}, a carbon-friendly ML inference service runtime system that balances performance, accuracy, and carbon emissions through mixed-quality models and GPU resource partitioning. Our experimental results demonstrate that \sol{} is effective in substantially reducing carbon emissions while maintaining high accuracy and meeting service level agreement (SLA) targets. 
\end{abstract}
\settopmatter{printacmref=false}

\maketitle
\pagestyle{empty}

\section{Introduction}
\label{sec:intro}


Reducing carbon emissions is of critical importance to combat the growing threat of climate change, as noted by United Nation and other agencies~\cite{un_report}. The large-scale systems that host information technology services account for 2\% of the global carbon emission~\cite{gupta2022act} -- the amount of datacenter workload has grown by 260\% in the past 6 years and is expected to keep growing, and, its contribution to the global carbon emission is likely to increase by many folds~\cite{iea_report,andrae2015global}. 

A relevant and contributing trend is that technology companies are increasingly incorporating artificial intelligence (AI) into their products and hosting trained machine learning (ML) models in datacenters GPUs to offer ML inference services to customers. Inadvertently, these inference services have exacerbated the carbon emission challenge because they account for a large proportion of the datacenter compute cycles. For example, many of Google's billion-user services are empowered by AI and their inference represents 60\% of the AI infrastructure emissions~\cite{patterson2022carbon}; Meta has expanded their infrastructure capacity by 2.5$\times$ to meet the ML inference demand~\cite{wu2022sustainable}; AWS and NVIDIA have estimated that inference accounts for 90\% of the ML workloads in HPC and cloud datacenters~\cite{aws_inf,nvidia_inf}.

\textit{Therefore, the goal of this paper is to design a novel carbon-friendly ML inference service runtime system.} Unfortunately, carbon-friendliness is often at odds with other desirable properties, including performance (low inference latency) and inference accuracy (high accuracy requires complex models and more computation-intensive operations, increasing the carbon footprint). But currently, we do not have the tools to effectively and automatically navigate this trade-off space and make ML inference services carbon-friendly. Nevertheless, given the growing importance of carbon-free operation in HPC systems and datacenters, finding solutions to this problem is becoming increasingly critical~\cite{nrel, google_carbonfree, microsoft_carbonfree, gocarbonfree247, li2023sustainable}. 

\vspace{3mm}
\noindent\textbf{\sol{} Key Ideas and Contributions.} The following summarizes the key insights behind \sol{}\footnote{Clover is a green plant with often three-toothed leaflets. The green color symbolizes carbon friendliness, and three leaflets correspond to carbon savings, accuracy, and performance.}, the challenges in exploiting the observed opportunities, and \sol{}'s contributions.

\vspace{2mm}
\noindent\textbf{Opportunity Space of Mixed-Quality Models and GPU Partitioning for Carbon Saving.} This is the first study to present experimental evidence to demonstrate the opportunities and trade-offs in mixed-quality models and GPU partitioning for carbon savings. Our experiments reveal that creating a mixture of model variants (i.e., a mixture of low- and high-quality models) can result in significant carbon savings while maintaining a high level of accuracy. Additionally, GPU partitioning can also contribute to carbon reduction by optimizing resource utilization, although it can lead to increased latency and potential violations of SLA targets.  \textit{Unfortunately, navigating this three-dimensional trade-off space (accuracy, carbon emission, and latency/SLA targets) is challenging, especially in the context that the carbon intensity of the energy sources powering the datacenters varies over time (Sec.~\ref{sec:motiv})}.\vspace{2mm}

\noindent\textbf{Novel Carbon-Aware ML Inference Framework.} \sol{} presents the design and implementation of a novel carbon footprint-aware ML inference service that reduces carbon emissions, attains high accuracy, and meets SLA targets. 
The proposed framework incorporates two seemingly unrelated ideas: mixed-quality models and hardware support for GPU partitioning, to minimize carbon footprint. \textit{\sol{}'s intelligent GPU partition boosts the utilization efficiency and provides the opportunity for carbon emission savings, but it can negatively impact SLA -- using mixed-quality model variants, \sol{} mitigates that challenge as lower-quality models allow \sol{} to minimize SLA violations, while higher-quality models allow \sol{} to reach high accuracy.} \sol{}'s optimization engine dynamically adapts to the carbon intensity of the energy source powering the large-scale data centers to opportunistically achieve its carbon and accuracy goals, while meeting SLA targets.  \sol{}'s optimization process is completely online, does not require offline training data, and is practical (Sec.~\ref{sec:design}).\vspace{2mm}

\noindent {\textbf{Evaluation Results.}} Our extensive evaluation of the \sol{} system on real-world scenarios demonstrates its effectiveness in reducing carbon emissions during model inference, while still achieving high accuracy and meeting SLA constraints. We evaluate the system in representative production environments, using real-world carbon intensity traces, and ML models, including the BERT model for natural language processing, object detection, and image classification applications. Our real-system prototype demonstrates that \sol{} is within 5\% of the practically-infeasible Oracle technique (Sec.~\ref{sec:evaluation}). We anticipate \sol{} serves as a catalyst for the community to enhance and develop carbon-aware ML inference services.

\section{Background}
\label{sec:backg}

\noindent\textbf{Carbon Intensity and Carbon Emission.} Electric power plants generate carbon dioxide ($\text{CO}_2$) during power generation, and this generated power is fed to datacenters that provide computing-based services. These hosted services consume a significant amount of compute hours and contribute to the carbon emission of the datacenters. To measure the environmental impact of generated/consumed energy, carbon intensity is used as a metric, which measures the amount of ($\text{CO}_2$) emitted per unit of energy (e.g., $\text{gCO}_2$/kWh).  Essentially, the carbon intensity indicates the ``greenness''  of the generated energy. Depending upon the time and geographical location, the greenness of the energy mix to the datacenter can vary. 

In the context of a datacenter, the carbon emission/footprint refers to the amount of greenhouse gases (g$\text{CO}_2$) emitted directly or indirectly from the site, and varies depending upon the ``greenness''  of the energy source powering the datacenter. Carbon footprint is often categorized into embodied carbon footprint and operational carbon footprint. Embodied carbon footprint is incurred when building the datacenter and manufacturing the components, and the operational carbon footprint corresponds to running computational workloads and hosting online services. Previous works have shown that embodied carbon footprint can be non-negligible for smartphones~\cite{gupta2022act}, but the operational carbon footprint is the biggest contributor (often more than 10$\times$ the embodied) toward the overall carbon footprint of the datacenters because a datacenter's operational energy is much larger in magnitude compared to mobile devices and their lifetimes are also much longer~\cite{lifecycle,frontier}. Therefore, reducing operational carbon footprint is a strong focus of datacenter operators~\cite{shin2021revealing,hsieh2020utilization}. Note that lowering the operational carbon footprint can indirectly lower the embodied carbon footprint too. For example, the sharing of hardware resources reduces the operational carbon footprint by reducing the energy, and it also reduces the number of hardware resources -- lowering the required embodied carbon footprint for building the system.

In this paper, we focus on the ML inference services running in these datacenters which contribute to the operational carbon footprint of the datacenters because they tend to make up for 90\% of total AI compute cycles~\cite{nvidia_inf,aws_inf}. Throughout this work, the term ``carbon'' refers to the \textit{operational carbon} by default, and we use the terms ``carbon footprint'' and ``carbon emission'' interchangeably. The carbon footprint is calculated as the product of energy consumed and carbon intensity of the energy source ($\text{Carbon Footprint} = \text{Energy} \times \text{Carbon Intensity}$), same as that defined by other works~\cite{lindberg2021guide,lacoste2019quantifying}. Lowering energy consumption (e.g., sharing hardware resources, and improving the power usage effectiveness with better cooling systems) reduces the carbon footprint of the online service. Similarly, using lower carbon-intensity energy sources (e.g., solar plant energy) to power a datacenter also reduces the overall carbon footprint. 

\vspace{1.5mm}
\noindent\textbf{Model Architecture Family and Variants.} When deploying ML models for inference, the same model architecture can have a family of model variants with a different number of parameters and sizes~\cite{wu2019wider,lan2019albert,tan2019efficientnet}, which offer flexibility in balancing accuracy and resource needs. The number of parameters in a model architecture depends on several factors, including the number of layers, the size of each layer, and the type of connections between the layers. Depending on the target number of parameters and computation resources, production AutoML tools~\cite{wang2021flaml,li2020system} can generate different fine-tuned model variants for the same application, which may offer different levels of accuracy and resource needs. For example, Google has pre-trained 8 model variants for the state-of-the-art EfficientNet~\cite{tan2019efficientnet} family, labeled from EfficientNet-B0 to EfficientNet-B7, as the model size increases. These variants have different numbers of parameters and have been traditionally used for deployment in different environments (e.g., resource-constrained edge devices, strict latency requirements, etc.)~\cite{hadidi2019characterizing,chen2020deep,wan2020alert}. 

\vspace{1.5mm}
\noindent\textbf{GPU Partitioning using Multi-Instance GPU.} For GPU partitioning for resource sharing, we are using the Multi-Instance GPU (MIG)~\cite{nvidia-mig} as a vehicle for experimental demonstration and evaluation. This is because MIG capability allows multiple workloads to share the same physical GPU, improving utilization while providing error and performance isolation -- prior GPU partition and sharing techniques do not offer this (e.g., Multi-Process Service~\cite{nvidia-mps}).

\begin{figure}
    \centering
    \includegraphics[scale=0.47]{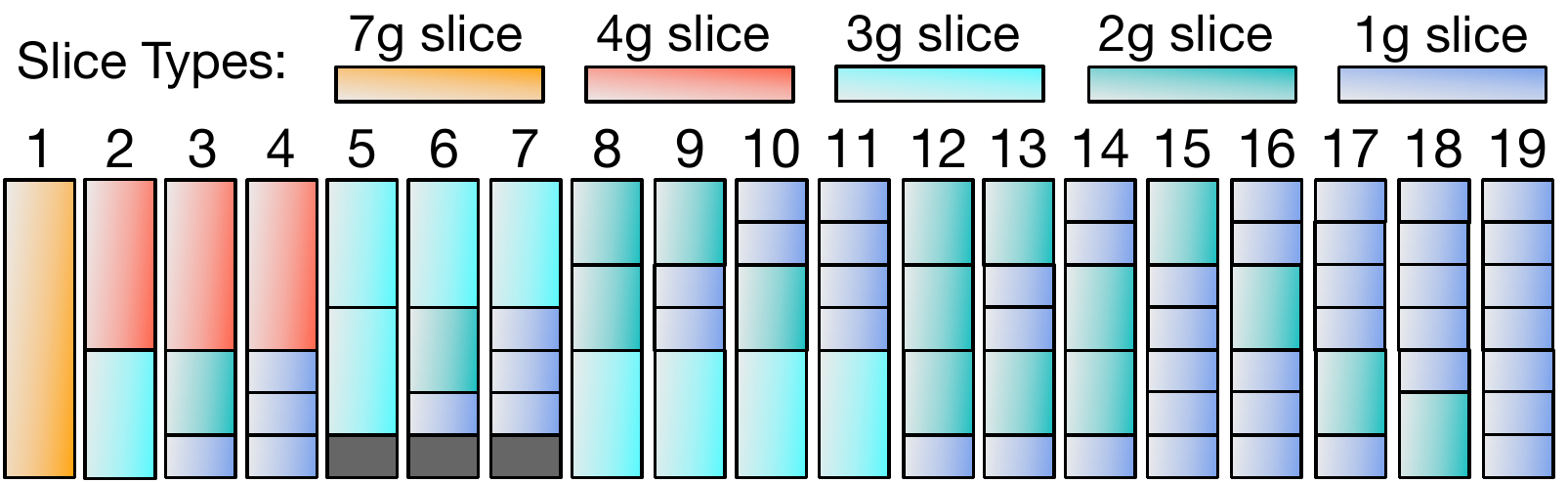}    
    \vspace{0.2cm}
    \hrule
    \vspace{-0.4cm}    
    \caption{Multi-instance GPU capability allows models to share a GPU via 19 different configurations. Each configuration is composed of a set of partitions. Each partition is one of the five different slice types representing different compute and memory capacities.}
    \vspace{-0.5cm}
    \label{fig:bkgd1_mig}
\end{figure}

Multi-Instance GPU (MIG)~\cite{nvidia-mig} is a hardware-based feature on NVIDIA Ampere, Hopper, and future generation datacenter GPUs to enable a single physical GPU to be partitioned into multiple smaller GPU slices, each with their own dedicated resources such as memory, compute cores, and cache. As shown in Fig.~\ref{fig:bkgd1_mig} (recreated based on~\cite{nvidia-mig}), there are a total of 5 MIG GPU slice types on NVIDIA A100: \texttt{7g}, \texttt{4g}, \texttt{3g}, \texttt{2g}, and \texttt{1g}, corresponding to the number of dedicated resources from highest to lowest (more details are available in ~\cite{li2022miso,li2022characterizing}). One can partition the GPU into 19 different MIG configurations consisting of these slice types. For example, when we set MIG configuration to configuration number 10, it partitions GPU into four slices of \{\texttt{1g}, \texttt{1g}, \texttt{2g}, \texttt{3g}\}. With the GPUs partitioned, one can host an instance of the inference service on every partition, increasing the total number of service instances compared to unpartitioned GPUs. Next, in Sec.~\ref{sec:motiv}, we provide details on how both the model variants and GPU partition interacts with the carbon footprint, accuracy, and latency of an inference service.










\section{Motivation}
\label{sec:motiv}

In this section, we provide three key insights that motivate the design of our \sol{} inference system. 

\begin{mybox}{}{}{\textbf{Opportunity 1.} Creating a mixture of model variants (i.e., a mixture of low- and high-quality models) provides an opportunity for a significant reduction 
in the carbon footprint while maintaining minimal loss in the overall inference accuracy.}
\end{mybox}

\begin{figure}
    \centering
    \includegraphics[scale=0.48]{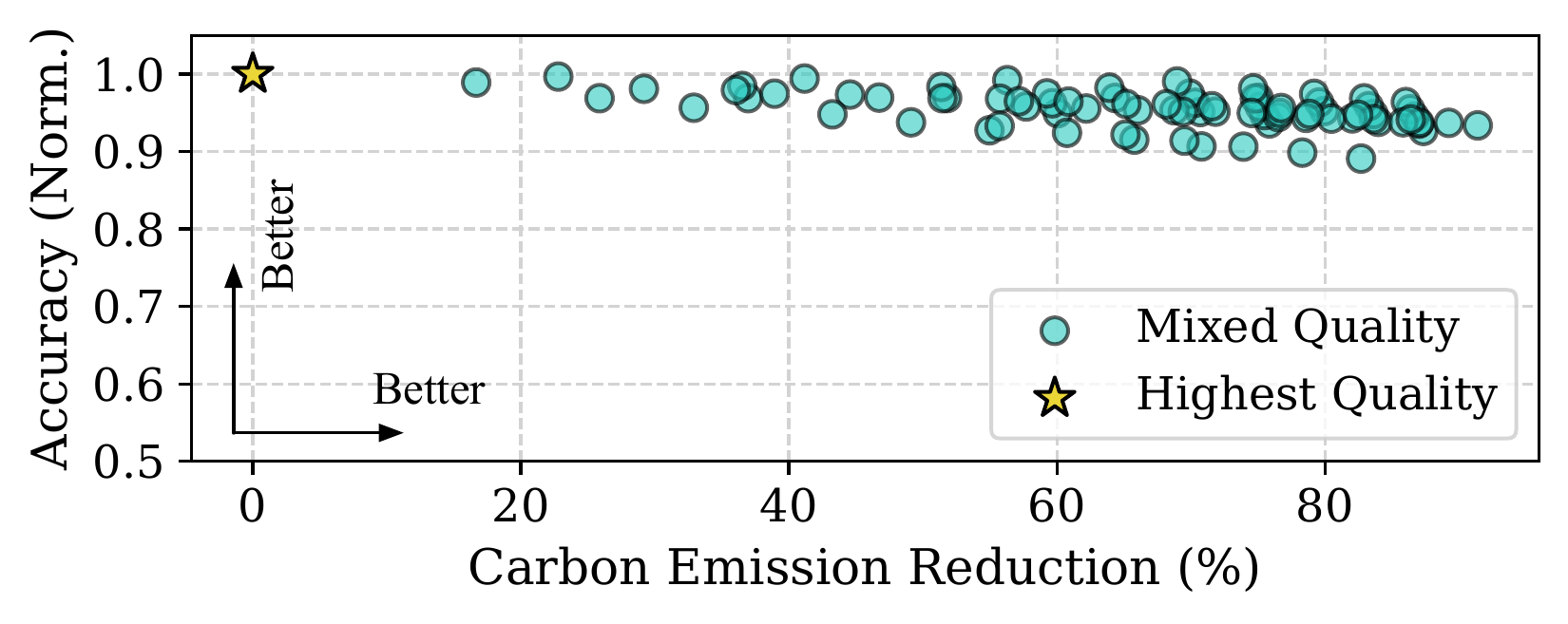}    
    \hrule
    \vspace{-0.4cm}    
    \caption{Using mixed-quality models provides an opportunity to reduce the carbon footprint of model serving significantly while maintaining high accuracy.}
    \vspace{-0.55cm}
    \label{fig:motiv1_variant}
\end{figure}

Fig.~\ref{fig:motiv1_variant} shows a visualization of using mixed-quality inference services hosted on a 4-GPU system (each GPU can host a different model variant). The x-axis represents the carbon footprint reduction and the y-axis represents the overall inference accuracy of different combinations of model variants (more methodology details available in Sec.~\ref{sec:eval_metho}). The overall accuracy is the weighted average accuracy of requests served by each model variant. Recall that different model variants can offer different levels of accuracy, and their corresponding carbon footprint also varies. A model that offers higher accuracy is likely to be more complex and more compute-intensive -- leading to higher operational energy and carbon footprint. 

The accuracy and carbon emission are relative to the default configuration where we host the highest-quality model variant on all GPUs, represented by the star point (0,1). For fair evaluation, while calculating the carbon emission/footprint savings, we have kept the carbon intensity (i.e., the greenness of the energy source powering the infrastructure) the same. Fig.~\ref{fig:motiv1_variant} reveals the trade-off between accuracy and carbon footprint reduction. For example, a combination of high-quality and low-quality models can provide over 60\% carbon footprint savings, while incurring less than 5\% accuracy degradation compared to the highest-quality model. As expected, the opportunity for carbon savings increases with a higher tolerance for accuracy degradation (more than 80\% carbon savings for 10\% accuracy loss). 

While model variants have been traditionally used by prior works for deploying in different environments (e.g., resource-constrained edge devices, strict latency requirements, etc.)~\cite{hadidi2019characterizing,chen2020deep,wan2020alert}. To the best of our knowledge, they have not been exploited for carbon savings because determining the combination of models with different accuracy/quality to attain a given accuracy degradation and carbon saving is challenging. Multiple models with different qualities may provide different carbon savings.  This problem becomes even more challenging in \sol{}'s case because we dynamically respond to the varying carbon intensity of the energy source (e.g., the fraction of energy coming from renewable sources may vary over the day) -- \sol{} solves these challenges. 

Next, we discuss the second aspect of the design trade-off -- latency of the ML inference requests. In particular, we show that GPU sharing among inference service instances can reduce carbon footprint, but increase the latency (carbon vs. latency trade-offs). 

\begin{mybox}{}{}{\textbf{Opportunity 2.} Partitioning a GPU hardware provides an opportunity to reduce the carbon footprint because it allows more efficient usage of GPU resources. However, sharing GPU resources leads to higher latency (i.e., higher SLA violation rate) -- making it challenging to exploit this opportunity space.}
\end{mybox}

\begin{figure}
    \centering
    \includegraphics[scale=0.46]{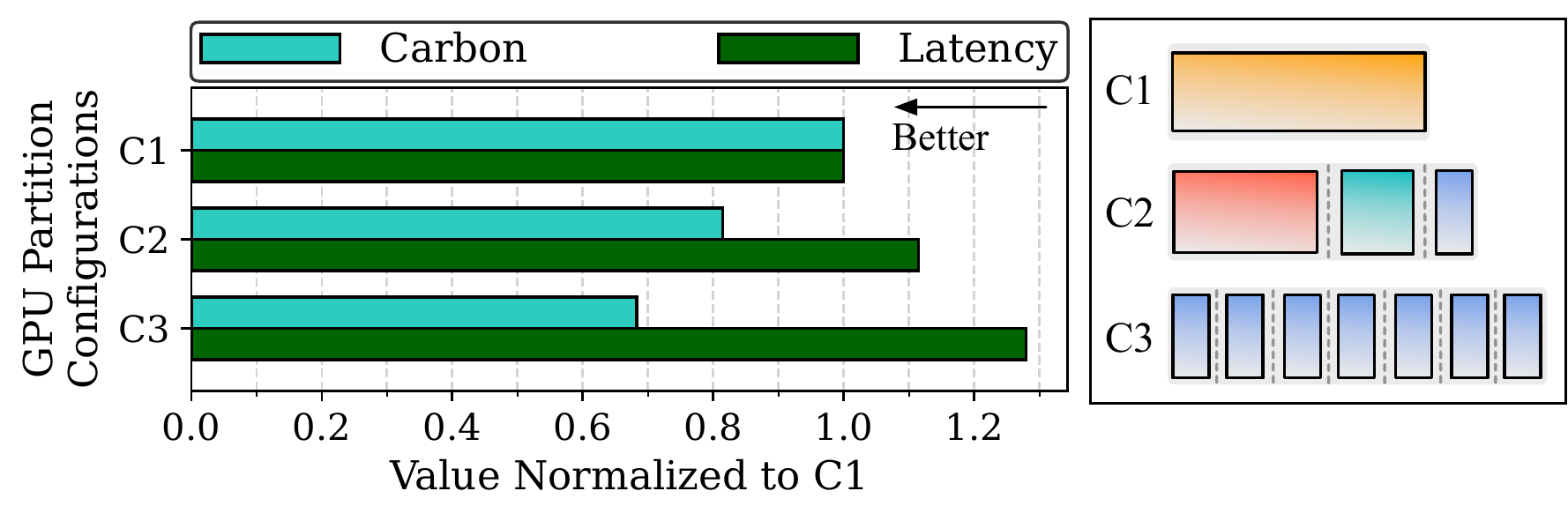}    
    \hrule
    \vspace{-0.4cm}    
    \caption{MIG GPU partitioning enables the opportunity to reduce carbon footprint because of better resource utilization, but at the cost of latency degradation. C1 is full GPU, C2 partitions the GPU into \{\texttt{4g}, \texttt{2g}, \texttt{1g}\} and C3 partitions the GPU into seven \texttt{1g} slices.}
    \vspace{-0.5cm}
    \label{fig:motiv2_mig}
\end{figure}

Fig.~\ref{fig:motiv2_mig} shows the effects of partitioning the GPU on the carbon footprint and the inference latency. Before drawing observations from this experiment, we briefly describe the setup. 

The GPU is configured to C1, C2, and C3, representing the MIG configurations 1, 3, and 19, respectively in Fig.~\ref{fig:bkgd1_mig}. We have hosted the same model variant (i.e., the model quality is fixed) across all partitions of the GPU and each partition hosts one instance of the inference service.  We keep the model quality the same to avoid complexities arising from accuracy trade-offs in this experiment -- which was the focus of the previous observation. For consistency, we keep the carbon intensity the same throughout the experiment. 

Fig.~\ref{fig:motiv2_mig} shows that when the GPU goes from a non-partitioned state (i.e., $C1$) to a fine-grained partitioned state ($C3$), we can reduce the carbon footprint by 30\% when serving the same number of inference requests. This is because fine-grained partitioning allows a higher degree of hardware sharing, and hence, better resource utilization. This leads to lower carbon emissions per request. 

All resources of a large, non-partitioned GPU can not be fully utilized by a single hosted model. Increasing sharing among multiple model instances allows better resource efficiency and hence, lower energy consumption (carbon footprint) per request on average. But, as expected, less dedicated resources from sharing hurts the latency of individual service instances. This implies that if we explore the option of GPU partition to reduce carbon footprint, this will inevitably violate the service level agreement (SLA) due to the latency increase. However, recall that different variants in a model family have different numbers of floating point operations during inference, hence different inference latency. While previous works characterized and utilized GPU sharing for different purposes~\cite{li2022miso,dhakal2020gslice,narayanan2020heterogeneity}, \sol{} is the first work to demonstrate that with carefully created mixtures of the model variants and GPU partition, we can reduce carbon footprint while meeting the SLA. 

In our experimental setups so far, we have kept the carbon intensity constant for a more accessible discussion of the trade-off among accuracy, carbon savings, and SLA. But, in practice, the carbon intensity of the energy source varies -- making the trade-off space even more challenging to exploit. 

\begin{mybox}{}{}{\textbf{Opportunity 3.} Carbon intensity significantly varies over time during different seasons and across geographical locations. An intelligent inference scheduler must exploit these spatial and temporal opportunities to navigate the trade-offs among carbon footprint savings, accuracy, and SLA. This can be achieved by dynamically creating a mixture of model variants and adjusting the GPU sharing among the model variants.}
\end{mybox}

\begin{figure}
    \centering
    \includegraphics[scale=0.5]{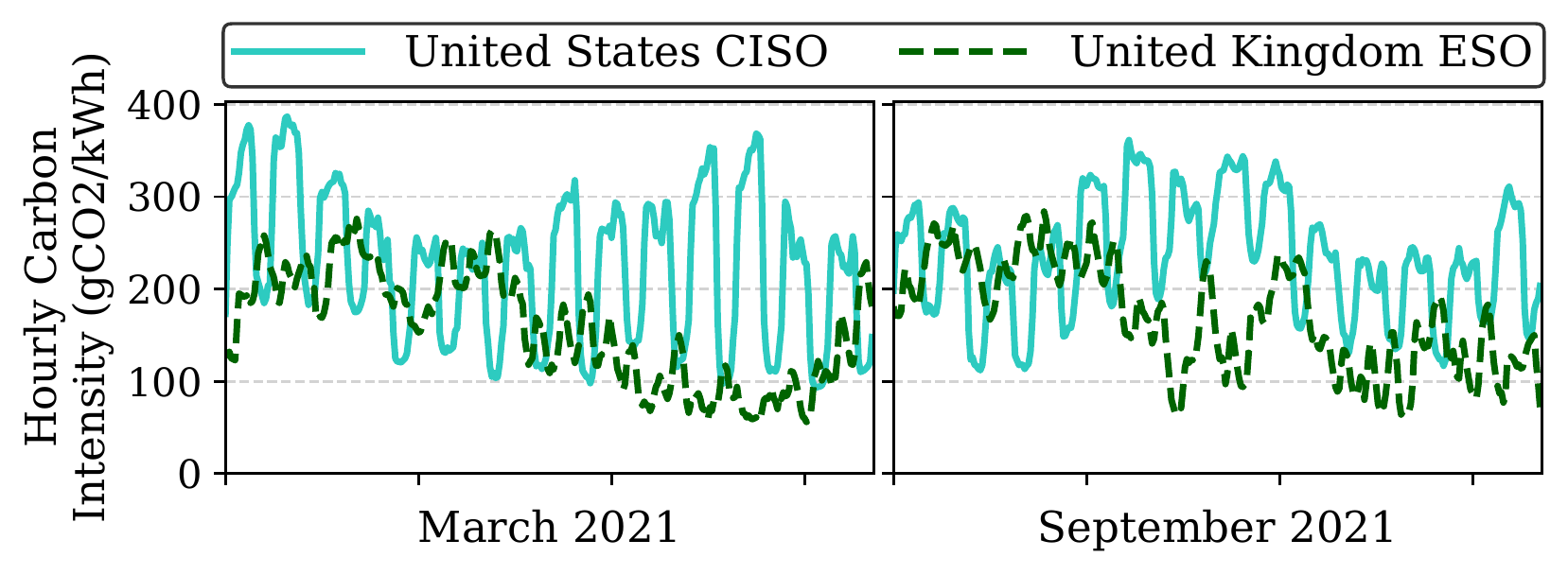}    
    \hrule
    \vspace{-0.4cm}    
    \caption{The carbon intensity during a 14-day span in March and September 2021 from two grid operators: California Independent System Operator (CISO) and UK Electricity System Operator (ESO).}
    \vspace{-0.5cm}
    \label{fig:motiv3_ci}
\end{figure}

Fig.~\ref{fig:motiv3_ci} shows the carbon intensity data that we collect from real power system operators in California and the UK~\cite{us-ci, uk-ci}. We observe that the carbon intensity can vary by more than 200 $\text{gCO}_2$/kWh within half a day, indicating that a carbon-aware inference system must continuously optimize its configuration based on the carbon intensity. The presence of variation in carbon intensity across different geographical locations and during different seasons reiterates the need for a carbon-aware solution. Fig.~\ref{fig:motiv3_ci} reveals that the patterns of the variation are distinctively different between regions over time due to different availability of unpredictable renewable energy sources, so fine-tuned solutions (e.g., adjusting inference configuration based on the exact time of the day and date of the year) for a particular time period is neither future-proof and effective in other geographical regions. Motivated by these trends, we present the design of \sol{}, the first carbon-aware machine learning inference system that exploits mixed-quality models and GPU partitioning to save carbon. \sol{} adapts to dynamic carbon intensity change to intelligently perform trade-offs among accuracy, carbon footprint, and SLA.

\section{\sol{} Design}
\label{sec:design}

\begin{figure}
    \centering
    \includegraphics[scale=0.361]{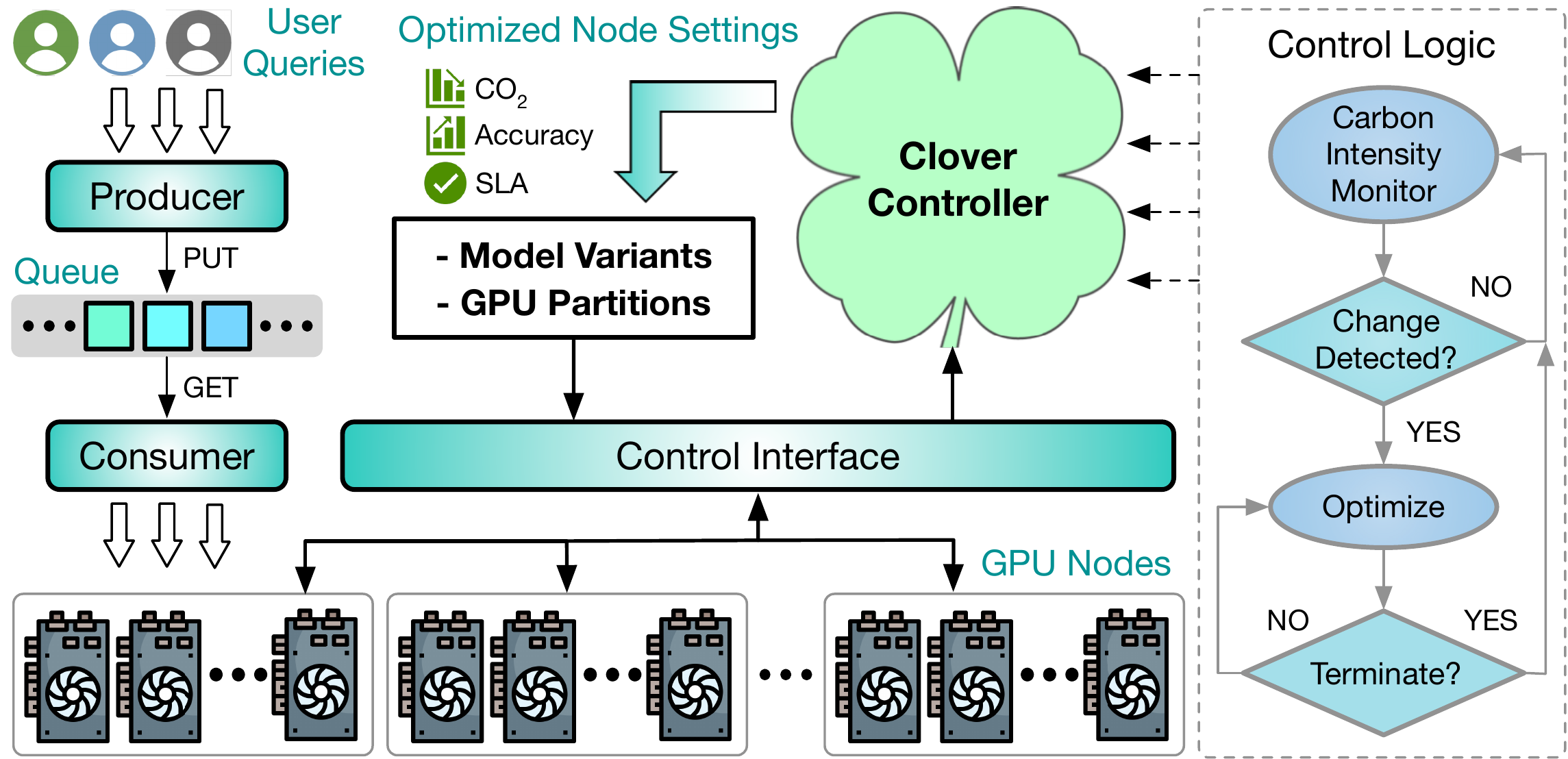}   
    \vspace{.1cm}
    \hrule
    \vspace{-0.4cm}    
    \caption{High-level system overview of \sol{}.}
    \vspace{-2mm}
    \label{fig:desi1_system}
\end{figure}

\sol{} is a carbon-aware system for hosting machine learning inference services on GPU servers. \sol{} intelligently creates a mixture of model variants on carefully selected GPU partitions to optimize for carbon footprint and inference accuracy under SLA constraints. Fig.~\ref{fig:desi1_system} visually depicts the system overview of \sol{}. The inference service accepts user queries online, these queries are put into a request queue by a producer module, and distributed by a consumer module to individual service instances hosted on the GPUs. The \sol{} controller is responsible for monitoring the real-time carbon intensity from the local grid and initiating its optimization process as a reaction to changes in carbon intensity or target accuracy threshold, etc. The optimization process generates new configurations of model variants and GPU partitions to configure the GPU nodes. For the rest of this section, we will explain: (a) How we formulate the carbon-aware inference challenge as an optimization problem; (b) How we devise a novel graph-based solution specific to the problem setup, and what happens during \sol{}'s optimization process; (c) Implementation details of \sol{}.

\subsection{Formulation of Carbon-Aware Inference}
\label{sec:desi_formulate}

\sol{} needs to optimize components simultaneously: a mixture of model variants (multiple low-quality and high-quality models), and the MIG partition of each GPU. By optimizing these components, we expect the inference service of \sol{} to provide higher inference accuracy when the carbon intensity is low to exploit the availability of renewable energy, while more aggressively reducing the relative carbon footprint when the carbon intensity is high. Throughout the process, \sol{} ensures that the users' quality of service is not compromised by strictly complying with the SLA (e.g., meeting the p95 tail latency requirement). 

\vspace{1.5mm}
\noindent\textbf{Optimization Variables.} Suppose there are $n$ GPUs allocated for the inference service. \sol{} needs to partition the GPU and create a mixture of model variants (a combination of multiple low-quality and high-quality models). We first define the optimization variable for GPU partition as $\bm{x}^p = [x_1^{p}, x_2^{p}, ..., x_n^p]$, where the $x_i^p \in \{1,2,...,19\}$ (see Fig.~\ref{fig:bkgd1_mig}) denotes the MIG partition configuration for the $i$-th GPU among $n$ GPUs. Let $m$ be the total number of GPU partitions in the system when applying the $\bm{x}^p$ partition. Note that $n \leq m \leq 7n$ because each GPU can have at most 7 partitions (see configuration 19 in Fig.~\ref{fig:bkgd1_mig}). Then, we can host $m$ service instances where every partition hosts one model copy. Each partition can host a different model variant (i.e., different quality). To formulate the model variant selection, all the variants available are encoded into ordinal data. For example, when hosting EfficientNets for image classification service, we can use 1 to represent EfficientNet-B1, 2 to represent EfficientNet-B2, etc. Then, the optimization variable for model variant selection on each GPU partition can be represented as $\bm{x}^v = [x_1^{v}, x_2^{v}, ..., x_m^v]$, where $x_j^v$ represents the model variant that is hosted on the $j$-th partition.

\vspace{1.5mm}
\noindent\textbf{Optimization Objectives.} \sol{} considers both the inference accuracy improvement and the carbon emission reduction in its objective function. We first define the accuracy baseline as $A_{base}$, which is the accuracy while only using the highest quality model variants in the service. Let $A(\bm{x}^p, \bm{x}^v)$ be the function of the actual overall inference accuracy when the configurations $\bm{x}^p,\bm{x}^v$ are used. Then, the inference accuracy difference can be represented as:
\begin{align}
    \label{eq:desi_acc}
   \Delta Accuracy = \frac{A(\bm{x}^p, \bm{x}^v) - A_{base}}{A_{base}} \times 100\% \vspace{-1mm}
\end{align}

Note that the $\Delta Accuracy \leq 0$ because $A_{base}$ represents the highest possible accuracy. Similarly, for optimization convenience, \sol{} defines a baseline reference for carbon footprint reduction. Specifically, we set the baseline $C_{base}$ ($\text{gCO}_2$/request) to be the average energy consumption per inference request when the service hosts the highest-quality model on a dedicated GPU, multiplied by a baseline average carbon intensity. In reality, average carbon intensity can vary over time, but for optimization purposes, assuming a baseline allows \sol{} to assess relative improvement, without loss of generality. This baseline is configurable and, does not impact the solution quality.
Let $ci$ denote the current carbon intensity (not an optimization variable), and $E(\bm{x}^p, \bm{x}^v)\cdot ci$ be the function of the actual average carbon footprint (energy multiplied by $ci$) per request when the configurations $\bm{x}^p,\bm{x}^v$ are used with current carbon intensity $ci$. We can calculate the carbon reduction as the following. 

\begin{align}
    \label{eq:desi_carbon}
   \Delta \text{Carbon} = \frac{C_\text{base} - E(\bm{x}^p, \bm{x}^v)\cdot ci}{C_{base}} \times 100\% \vspace{-1mm}
\end{align}

\noindent\textbf{Problem Formulation.} Next, \sol{} combines the accuracy and carbon footprint into a single objective function $f$ to maximize: \begin{align} \label{eq:desi_obj} f(\bm{x}^p, \bm{x}^v) = \lambda \cdot\Delta Carbon + (1-\lambda)\cdot\Delta Accuracy\vspace{-1mm} \end{align} where $\lambda \in [0,1]$ is a configurable parameter that the inference service provider can set to dictate how important the carbon footprint is compared to the inference accuracy, which depends on the application scenario. Our evaluation demonstrates that \sol{} is effective across different values of this configurable $\lambda$ parameters and can even treat accuracy as a threshold. The final piece of the problem formulation is the SLA constraint. In \sol{}, we use the p95 tail latency target as SLA, the same as other online service works~\cite{gupta2020deeprecsys,ali2020batch,liu2022veltair}. We denote the latency target as $L_{tail}$, and the actual tail latency of the deployed service with configurations $\bm{x}^p,\bm{x}^v$ as $L(\bm{x}^p,\bm{x}^v)$, \sol{} must guarantee that $L(\bm{x}^p,\bm{x}^v) \leq L_{tail}$. Putting everything together, we can formulate the problem as: \begin{align} \label{eq:desi_opt} &\max_{\bm{x}^{p}, \bm{x}^{v}} \begin{aligned}[t] & f(\bm{x}^p, \bm{x}^v) \end{aligned} \\ & \label{eq:constraint}\;\;\text{s.t. } \;L(\bm{x}^p,\bm{x}^v) \leq L_{tail} \end{align} 

\begin{figure} 
    \centering 
    \includegraphics[scale=0.266]{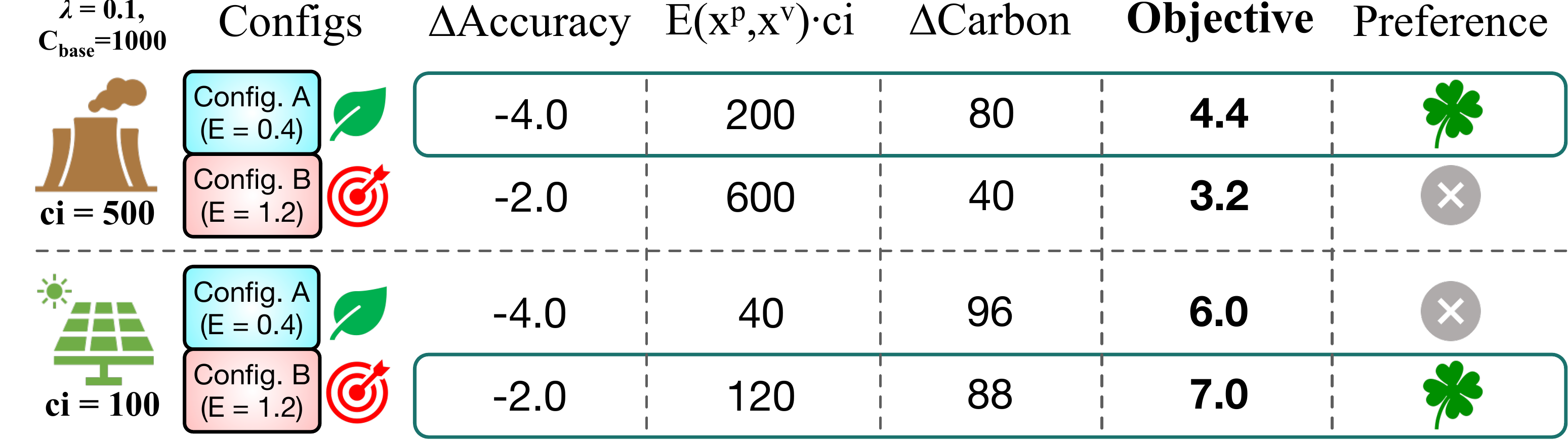} \vspace{.1cm} 
    \hrule 
    \vspace{-0.4cm} 
    \caption{\sol{} selects configurations based on carbon intensity.} 
    \vspace{-2mm} \label{fig:desi3_objective} 
\end{figure} 

This formulation empowers \sol{} to become carbon-aware when the environment's carbon intensity varies. We set up two illustrative example configurations $A$ and $B$ in Fig.~\ref{fig:desi3_objective}, where $A$ has a smaller carbon footprint while $B$ has higher prediction accuracy. Fig.~\ref{fig:desi3_objective} shows that \sol{} selects config. $A$ when the carbon intensity is high ($ci=500$). When the carbon intensity drops low ($ci=100$) at a later point, $\Delta Accuracy$ remains the same but $\Delta Carbon$ changes (Eq.~\ref{eq:desi_carbon}), therefore, \sol{} prefers config. $B$ (the higher quality service configuration). Next, we discuss how \sol{} solves this problem by formulating this as a graph-optimization problem.

\begin{figure*}[ht]
    \centering
    \includegraphics[scale=0.398]{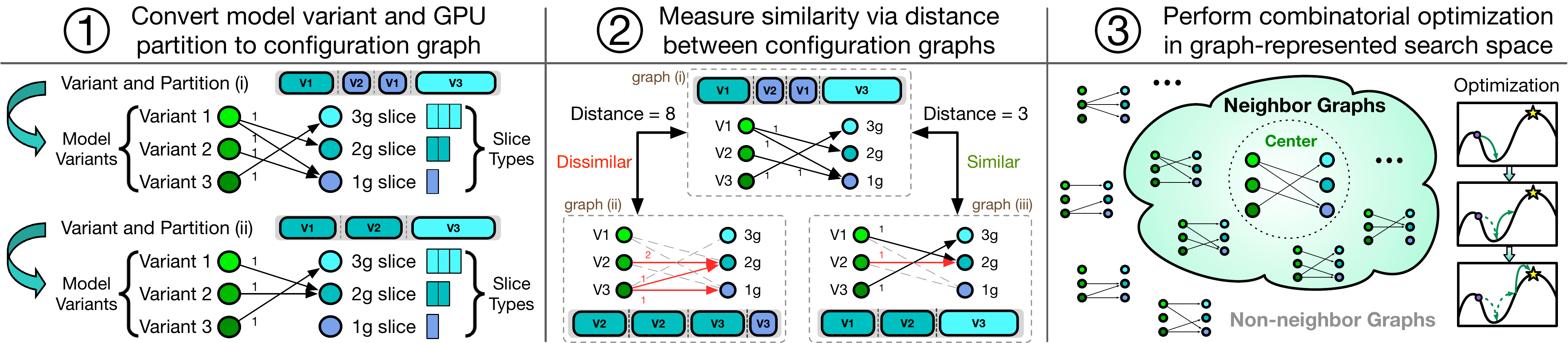}   
    \vspace{.1cm}
    \hrule  
    \vspace{-4mm}
    \caption{Key steps of \sol{} configuration-graph-based optimization.}
    \vspace{-4mm}
    \label{fig:desi2_optimize}
\end{figure*}

\subsection{Configuration-Graph-Based Optimization}
\label{sec:desi_graph}

\sol{} needs to solve a combinatorial optimization problem since the $\bm{x}^p,\bm{x}^v$ variables take discrete values from a large configuration space. This is because $\bm{x}^p$ is n-dimensional, and $\bm{x}^v$ is m-dimensional where $m=O(n)$. \textit{Unfortunately, the problem cannot be solved effectively using analytical methods, as we cannot derive the exact mathematical expression for functions $f$ and $L$. One potential solution is to train a large neural network to model the interaction between the configuration and the relevant metrics. However, this has limited practical effectiveness due to the offline training overhead, limited portability across scenarios, and the need for a large amount of training data.} Therefore, \sol{} devises an \textit{online} approach that does not require training data, works across scenarios, and is effective. 

Before presenting the \sol{}'s optimizing process, we first make an observation that \sol{}'s optimization variables can be mapped to a graph representation form (discussed next). Therefore, \sol{} leverages the graph similarity theory to solve this optimization problem by applying simulated annealing in the graph space.

\vspace{1.5mm}
\noindent\textbf{\sol{} Configuration Graph. } Fig.~\ref{fig:desi2_optimize} visually depicts \sol{}'s optimization process. \sol{} maps the optimization variable $(\bm{x}^p, \bm{x}^v)$ into a configuration graph representation denoted as $x^g$.  The \sol{} configuration graph is formally defined as:

\begin{definition}
A \sol{} configuration graph is a directed bipartite graph consisting of two types of vertices: the \textbf{model variant vertices} and the \textbf{MIG slice vertices}. Each model variant vertex (or variant vertex, for short) represents one unique model variant from the model architecture family (Sec.~\ref{sec:backg}); each slice vertex represents one unique MIG slice type available from the GPU (Sec.~\ref{sec:backg}). A weighted edge connects a variant vertex to a slice vertex, where the weight represents the total number of the particular model variant instances hosted on the corresponding MIG slice type. 
\end{definition}


In Fig.~\ref{fig:desi2_optimize} step 1, we show two different $(\bm{x}^p, \bm{x}^v)$ configurations on one GPU and simplify the graph to three vertices of each vertex type to simplify the demonstration. In practice, there are five slice vertices for NVIDIA A100 and H100 GPUs (Fig.~\ref{fig:bkgd1_mig}), and the number of variant vertices equals the number of model variants that the service provider develops. The total weight of all edges equals the total number of service instances or model copies. Recall that a configuration can have multiple slices of the same type, and this is why the edge can have weight $>1$, even though each slice/partition holds only one model inference instance. 

\vspace{1.5mm}
\noindent\textbf{Why optimize using a graph-based representation?} \sol{}'s choice of representing the optimization variable as a configuration graph has two major benefits. Firstly, a graph representation allows for the removal of configurations that yield the same objective function values, resulting in a more compact search space. As discussed in Sec.~\ref{sec:backg}, MIG provides performance isolation between workloads on the same GPU, meaning that only the MIG slice type that the model copy runs on is relevant. This slice type configuration is represented as the edges in the configuration graph. Which GPU the copy runs on, and which model variants it is sharing the same GPU with may result in different $(\bm{x}^p, \bm{x}^v)$ values, but they all result in the same objective function value and the same graph $x^g$. 

The second advantage is that the configuration $x^g$ enables better additivity, meaning that when we add more GPUs to the system, we can simply add the edge weights of the configuration graph for the new GPUs to the current graph $x_g$. Similarly, we can also do edge weight deduction when removing GPUs. Because of this additivity property, we can easily modify the edge weight values when varying the number of GPUs, and the number of vertices remains constant. In contrast, the original representation requires changing the dimensionality of $\bm{x}^p$ and $\bm{x}^v$ when the GPU number changes. We note that although the smallest MIG slice on 40GB NVIDIA A100s has sufficiently large memory (5\,GB), not all models can always fit in the smaller GPU slices. 
As such, \sol{} has taken this into account appropriately by disabling the edge connection between corresponding variant and slice vertices if out-of-memory errors would occur.

\vspace{1.5mm}

\noindent\textbf{Graph Similarity.} Once \sol{} has the graph space representation of the configuration, we need a metric to determine how similar or dissimilar different configurations are. This is because solving combinatorial optimization problems involves exploitation and exploration, which is more effective if the similarity of the samples can be established.
To address this challenge, \sol{} uses the graph edit distance (GED)~\cite{sanfeliu1983distance} to measure the similarity between two configuration graphs, which is a widely used metric in pattern recognition~\cite{gao2010survey,papadopoulos2022learning,jiao2022sequential}. Since our graphs contain weighted edges and all configurations have the same vertices, we consider the edge weight in computing the GED. For example, we show two examples of GED computation in Fig.~\ref{fig:desi2_optimize} step 2. For the comparison from graph (i) to (ii), the editing includes removing all current edges of weight 1, and adding two new edges of weight 1 and one edge of weight 2, the GED is computed as 8. A shorter distance means higher similarity, thus graph (iii) is considered more similar to (i) than graph (ii). This similarity measurement also makes sense from the variant and partition point of view as we show the actual configuration behind the graph representation in Fig.~\ref{fig:desi2_optimize} step 2.

\vspace{1.5mm}
\noindent\textbf{Optimization in the Graph Space.} Fig.~\ref{fig:desi2_optimize} step 3 visually demonstrates how \sol{} performs optimization using the graph-represented variables. First, we define the neighbor configurations: marking a particular configuration as a center, all other configurations whose graph edit distance (GED) to the center configuration is within a distance threshold are considered as neighbors of the center. \sol{} sets this GED threshold to be four because swapping the model variant of one service instance incurs two GED and switching a model copy to be hosted on a different MIG slice type also incurs two GED. 

Now that \sol{} has identified similar and dissimilar configurations in the graph, it applies combinatorial optimization algorithms in the graph-represented search space. \sol{} uses the simulated annealing (SA) algorithm~\cite{kirkpatrick1983optimization}, which is a simple and effective optimization method ~\cite{bianco2018polymer,mirhoseini2021graph,wang2015parallel}. \sol{} follows the basic principles of simulated annealing: during optimization, it randomly samples a new graph ${x^g}'$ from the neighborhood of the current center $x^g$ to evaluate. If ${x^g}'$ is a lower energy solution, we accept the new configuration and move the center to ${x^g}'$ so that we can sample within the new neighborhood. Note that ``energy'' is an SA-specific optimization function that SA minimizes, and should not be confused with the energy consumption of GPU. We design the SA algorithm's energy function to incorporate both \sol{}'s objective function and the SLA as $h$.
\begin{align}
\label{eq:desi_sa1}
    h(x^g) = -f(x^g)\cdot \min\{1, \frac{L_{tail}}{L({x^g})}\}
\end{align}

Since SA minimizes energy $h$, we negate the objective function $f$. The $\min$ operator represents that when the configuration meets SLA, we only care about $f$. But if $L(x^g) > L_{tail}$, \sol{} punishes the function by how much it violated the SLA to maintain the smoothness of the space. We accept 
${x^g}'$ if $h({x^g}') \leq h(x^g)$. Otherwise, we accept ${x^g}'$ based on a probability function $P(x^g, {x^g}')$ which follows the standard form of SA algorithms:
\begin{align}
\label{eq:desi_sa2}
    P(x^g, {x^g}') = exp({-\frac{h({x^g}') - h(x^g)}{T}})
\end{align}
Here, $T$ is the SA temperature parameter (and not the temperature of the GPU). We set $T=1$ initially and use a cooling schedule of $0.05$ per iteration until the temperature drops to 0.1, which means the probability of accepting a worse configuration is higher early to escape local optima and gets lower as we evaluate more samples. We terminate the optimization process once it reaches a 5-minute time limit or no better configuration has been found in 5 consecutive evaluations. \sol{} re-invokes the optimization using a service provided pre-determined configurable parameter -- for example, change in the carbon intensity by more than 5\%, violation in the accuracy threshold, change in the SLA limit, or change in the $\lambda$ parameter that represents the relative weight between the carbon emission and the inference accuracy.

In summary, \sol{} formulates the challenges of carbon-aware inference as an optimization problem and designs a novel graph-based optimization technique to optimize the dual objective of accuracy and carbon footprint while guaranteeing the SLA.

\subsection{\sol{} Implementation}
\label{sec:desi_implement}

\sol{} framework is implemented as a combination of three service modules: (a) the load balancer module, (b) the GPU node service module, and (c) the master controller module. 

The load balancer module runs in a separate CPU-based server, a producer accepts user requests and put them in a FIFO queue which is monitored by a consumer (Fig.~\ref{fig:desi1_system}). Whenever a service instance finishes processing, it notifies the consumer, which removes the request from the head of the queue and feeds it to the service instance. If the consumer cannot keep up with the producer due to a bad configuration or limited capacity, requests will spend excessive time waiting in the queue, leading to SLA violations. Therefore, \sol{}'s SLA constraint implicitly posts a throughput constraint to the service. 

The GPU node service module consists of workers that process the inference requests. The inference services are hosted as HTTP endpoints using the Python Flask library. When the GPU gets partitioned, each MIG slice is assigned a unique CUDA device ID. We set the \texttt{CUDA\_VISIBLE\_DEVICES} environment variable to the device ID to host a model on a specific partition. In addition to the inference service, each GPU node also hosts a service that communicates with the master controller node and a carbon measurement service. We develop the carbon measurement service based on carbontracker~\cite{anthony2020carbontracker}, a Python library that measures the total energy consumption of the CPU and GPU components then converts it to the carbon emission. We modify carbontracker to interact with the master controller and support model inference.

\sol{}'s master controller runs on a separate CPU node and is responsible for all the behind-the-scene optimizations. During the optimization, it collects the accuracy, carbon, and latency data, runs the SA algorithm, determines the next configuration to evaluate, and reconfigures the GPU nodes. \sol{}'s optimization runs with graph-represented GPU node configurations, which is implemented using the Python NetworkX library~\cite{hagberg2008exploring,SciPyProceedings} - the overhead of running optimization in the background is included in all our results, including the time taken to re-partition the hardware and reinitialize the new service instances.  We note that ML inference services can be hosted in a modular fashion (a pool of pods with multiple A100 in each pod). Hence, a hierarchical approach or grouping-based scalability can be attained in practice by \sol{}. \sol{} runs as a master controller on a CPU node separate from the worker GPU nodes, using one CPU core at near-100\% utilization, which adds less 0.5\% critical path latency and energy overhead, both included in reported results.

\section{\sol{}: Experimental Evaluation}
\label{sec:evaluation}

\subsection{\sol{} Methodology}
\label{sec:eval_metho}

\begin{table}
\centering
\caption{Machine learning inference applications}
\vspace{-4mm}
\scalebox{0.76}{
\begin{tabular}{cccc}
\toprule
\textbf{Application} & \textbf{Dataset} & \textbf{Architecture} & \textbf{Variants} \\ 
\midrule
\midrule
\makecell[c]{Object\\Detection} & \makecell[c]{MS COCO~\cite{lin2014microsoft}\\(Microsoft)}  & \makecell[c]{YOLOv5~\cite{redmon2016you}\\(Ultralytics)}  & \makecell[c]{YOLOv5l, YOLOv5x,\\YOLOv5x6} \\
\midrule
\makecell[c]{Language\\Modeling} & \makecell[c]{SQuADv2~\cite{rajpurkar2018know}\\(Stanford)} & \makecell[c]{ALBERT~\cite{lan2019albert}\\(Google)} & \makecell[c]{V2-base, V2-large,\\V2-xlarge, V2-xxlarge}  \\
\midrule
\makecell[c]{Image\\Classification} & \makecell[c]{ImageNet~\cite{russakovsky2015imagenet}\\(Princeton/Stanford)} & \makecell[c]{EfficientNet~\cite{tan2019efficientnet}\\(Google)} & B1, B3, B5, B7  \\

\bottomrule
\end{tabular}}
\vspace{-2mm}
\label{table:metho}
\end{table}

\noindent\textbf{Evaluation setup. } The \sol{} framework is implemented and experimentally evaluated on the testbed consisting of five compute nodes, each node has two NVIDIA A100 40GB Tensor Core GPUs with two AMD EPYC 7542 CPUs -- a total of ten NVIDIA A100 GPUs (195 TFLOPS), which is similar to the computational power of inference servers tested in MLPerf~\cite{reddi2020mlperf}, and more or similar to other works that use A100 GPUs, such as Abacus~\cite{cui2021enable} and FpgaNIC~\cite{wang2022fpganic}. Recall that each NVIDIA A100 GPU can have 7 MIG NVIDIA A100 slices each and hence, a total of 70 MIG slices are available for hosting our models. 

We evaluate \sol{} on three representative ML inference applications: object detection, language modeling, and image classification. These applications use pre-trained state-of-the-art ML models from the industry, and use validation sets of the most commonly used datasets for inference queries. The details are listed in Table~\ref{table:metho}. The accuracy numbers for model variants are available from the public repositories of the models~\cite{accuracyref1, accuracyref2, accuracyref3}, and our evaluation uses these numbers. However, the key concepts introduced in \sol{} are not restricted to these models and extend to other models with different quality variants. 


For each application, the default configuration is to host the largest model variant on the ten GPUs in our system without MIG partitioning. We model the user queries using Poisson distribution, following the standard methodology used in previous works of ML inference service~\cite{reddi2020mlperf,kannan2019grandslam,shen2019nexus}. We set the Poisson parameter such that there is neither resource starvation nor idle GPUs and set the SLA to be the p95 tail latency~\cite{gupta2020deeprecsys,ali2020batch,liu2022veltair} running the default configuration. For any combination of model variants and GPU partitions that \sol{} configures later, its p95 tail latency must be within the SLA target. We note that the p95 tail latency from the baseline (without MIG partitioning) case is not adjusted when \sol{} partitions the GPUs -- the same p95 tail latency from the base case is continued to be used as an SLA constraint which makes it realistic for \sol{} to operate in production environments.

We note that \sol{}'s evaluation purposely reports improvements for each ML inference model to demonstrate that \sol{}'s design is effective across different types of models individually. 
While \sol{} supports co-locating models of different types on the same GPU (e.g., language and object detection), datacenters may prefer a more practical approach, such as managing separate pods of servers, where each pod serves a specific model type. This approach helps avoid unpredictable performance and networking interference among different model types. 
\sol{} demonstrates that it is possible to save carbon emissions and meet SLAs for different types of models without requiring datacenter operators to change their current practices or unfairly penalize one application over others. Our aggregate savings represent the average of the three models, but we present them individually to confirm that each model can benefit from \sol{}'s optimizations.

\begin{figure}
    \centering
    \includegraphics[scale=0.49]{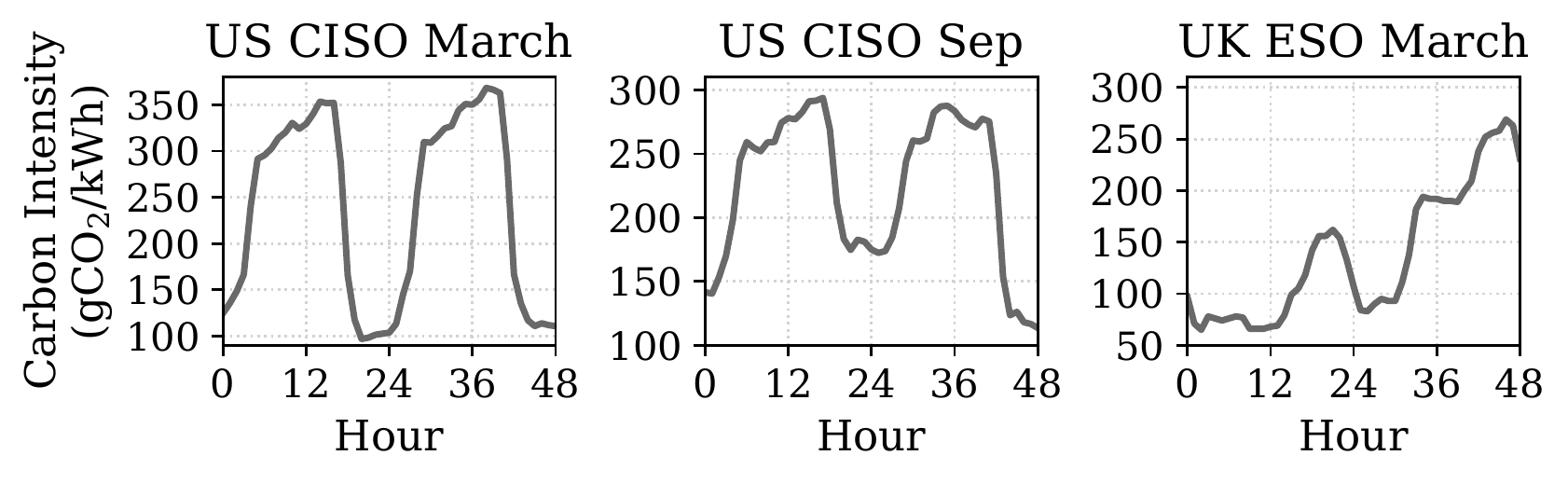}    
    \hrule
    \vspace{-0.4cm}    
    \caption{Representative carbon intensity traces used for evaluation.}
    \vspace{-2mm}
    \label{fig:eval0_ci}
\end{figure}

We use real-world carbon intensity traces to evaluate the effectiveness of \sol{}. To ensure the evaluation finishes within a reasonable amount of time while capturing enough variation in the carbon intensity, we set the trace span to be 48 hours, as shown in Fig.~\ref{fig:eval0_ci}. We use the US CISO (California) March trace throughout the evaluations in Sec.~\ref{sec:eval_eval}, and use the US CISO September and UK ESO March traces to demonstrate \sol{}'s effectiveness across geographical locations during different seasons. 
We note that while a datacenter's low power usage effectiveness (PUE) can contribute to carbon savings (e.g., better cooling infrastructure), \sol{}'s benefits do not depend on any assumptions about PUE. \sol{} is evaluated using a constant PUE value of 1.5~\cite{uptime}. But, the benefits are reported as relative to ensure its benefits are not dependent on PUE values. Although there is a lack of open-access real-time PUE data, the PUE optimizations are complementary to ideas presented in \sol{} and not competitive. The choice of PUE is based on prior work~\cite{uptime}, but other values can also be used. 

The $\lambda$ parameter which represents the relative weight between the carbon emission and the inference accuracy, is set to $0.5$ by default. \sol{} always runs multiple inference server processes simultaneously.

\vspace{1.5mm}
\noindent\textbf{Competing schemes.} We compare \sol{} to four other schemes: a baseline scheme (carbon-unaware, but highest accuracy), a carbon-optimal scheme, a carbon-aware system called \blover{} that is an inferior version of \sol{}, and an oracle scheme. To the best of our knowledge, \sol{} is the first carbon-aware ML inference system with complex optimization objectives. To establish an upper bound on benefits, we designed an oracle scheme that uses oracular knowledge for optimizing the same metrics as \sol{}. This also demonstrates that \sol{} is close to the upper bound that a hypothetical technique can achieve.

\textbf{BASE. } We use the highest-quality model that runs on every GPU exclusively as the \base{} scheme, equivalent to the default configuration. \base{} has the advantage that it always provides the highest inference quality with the lowest latency possible, but it does not explore the opportunities we discussed in Sec.~\ref{sec:motiv}.

\textbf{\opt{}. } This represents a scheme that aggressively minimizes \coo{} by using the most aggressive GPU partition (configuration 19 in Fig.~\ref{fig:bkgd1_mig}) and hosting the smallest model variant on each partition. \opt{} exploits the insights that we show in Sec.~\ref{sec:motiv}, and uses the smallest model variant to meet the SLA. However, it does not try to improve inference accuracy based on carbon intensity. 

\textbf{\blover{}. } \blover{} stands for Basic-\sol{}. This is a home-grown variant of \sol{} which leverages mixed-quality models and GPU partition and is carbon-aware just like \sol{}. \blover{} implements all of \sol{}'s design principles in Sec.~\ref{sec:design} except the graph-based optimization in Sec.~\ref{sec:desi_graph}. To optimize according to current carbon intensity, \blover{} performs random search in the original problem space defined by $\bm{x}^p$ and $\bm{x}^v$.
Therefore, \sol{}'s superiority over \blover{} specifically demonstrates the effectiveness of \sol{}'s graph-based optimization.

\textbf{\oracle{}. } This is an oracle scheme obtained from exhaustive offline profiling. Whenever the carbon intensity changes, \oracle{} instantly adjusts to the variant selection and GPU partition that maximizes the objective function. Although it guarantees optimality and is artificially designed to incur zero optimization time, it is infeasible to be deployable in practice. In our evaluation, we make at most four model variants available (Table~\ref{table:metho}) and standardize the model mixture and partition across all GPUs to limit the search space, but it still took the \oracle{} scheme approximately two weeks to complete its offline profiling, while \sol{} operates entirely online without futuristic information.

All of the above schemes except \base{} are designed to leverage different model variants and GPU partitions, which are the key insights as a part of \sol{}'s unique contributions. Next, we extensively evaluate \sol{} against the competing schemes to compare their inference accuracy and carbon footprint.

\subsection{\sol{} Evaluation and Analysis}
\label{sec:eval_eval}

\begin{figure}
    \centering
    \includegraphics[scale=0.515]{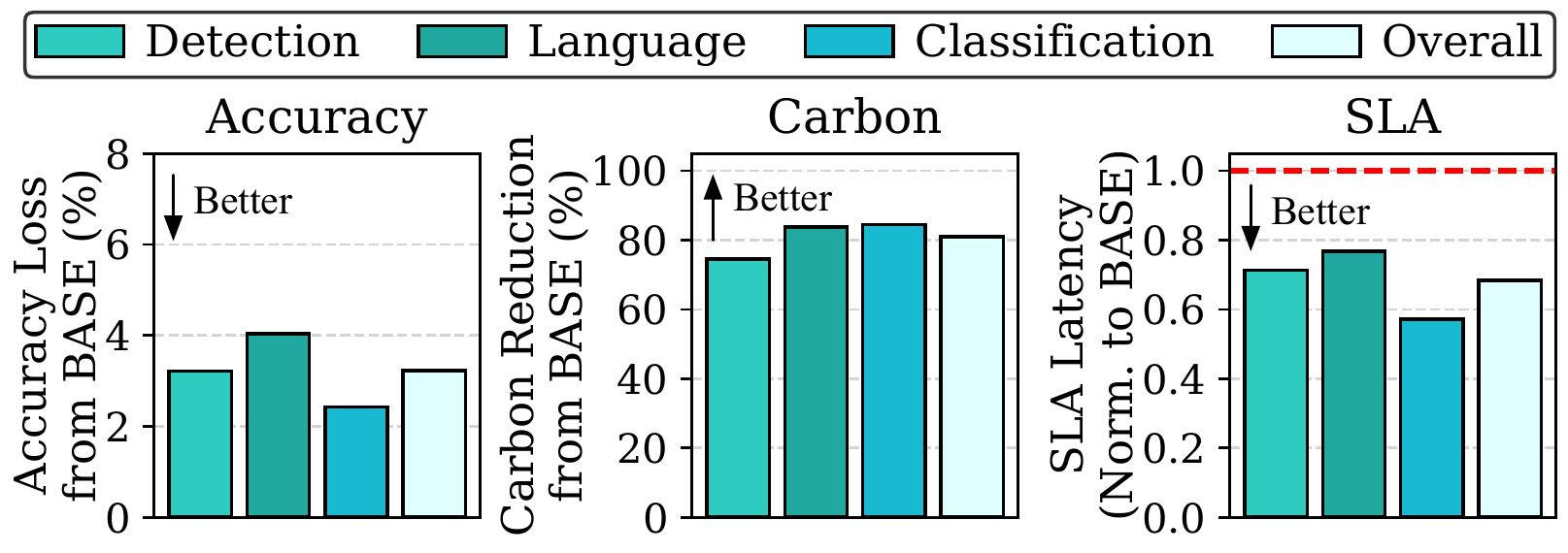}    
    \hrule
    \vspace{-0.4cm}    
    \caption{\sol{}'s effectiveness in accuracy, carbon footprint, and SLA (p95 tail) latency  compared to the baseline strategy.}
    \vspace{-0.5cm}
    \label{fig:eval1_clover}
\end{figure}

\subsubsection{Effectiveness of \sol{}}
\label{sec:eval_effective}

\begin{figure}
    \centering
    \includegraphics[scale=0.5]{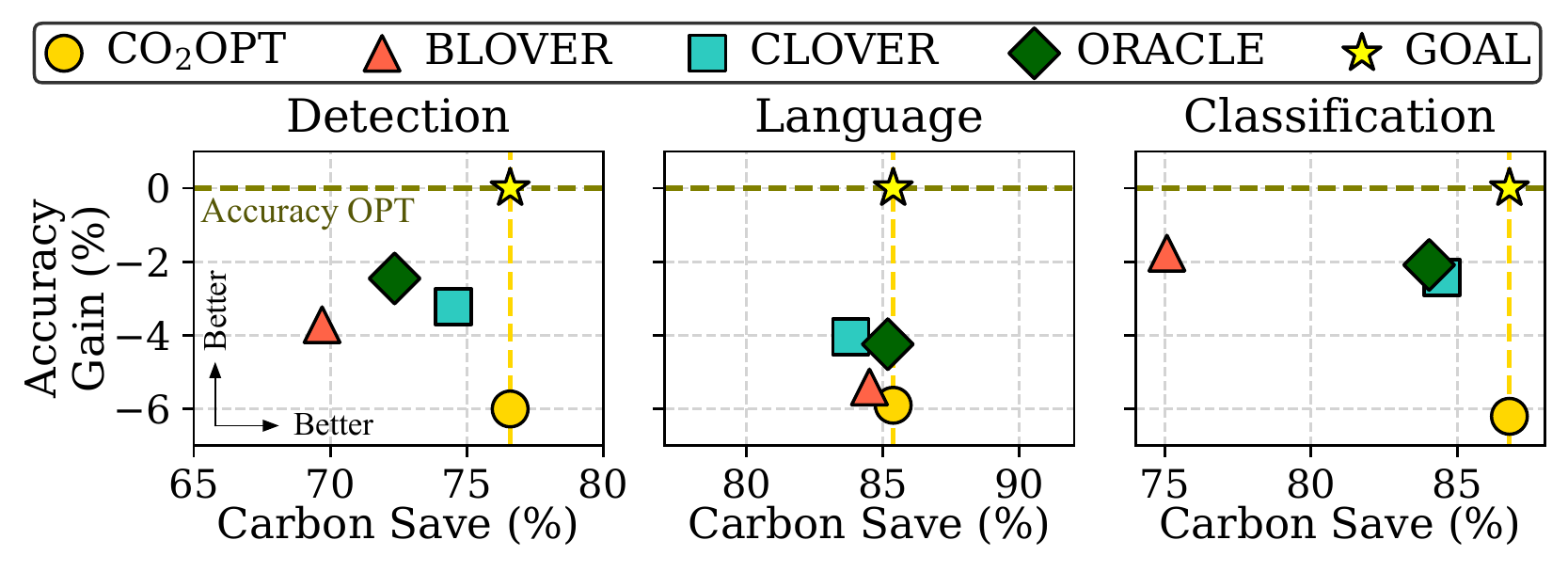}    
    \hrule
    \vspace{-0.4cm}    
    \caption{\sol{} effectively balances competing goals of carbon savings and accuracy. \sol{} provides nearly as much carbon saving as the carbon-optimal method, but with higher accuracy. \sol{} is the closest to \oracle{} and outperforms its basic version (\blover{}).}
    \vspace{-0.5cm}
    \label{fig:eval2_comnpare}
\end{figure}

\noindent\textbf{\sol{} significantly reduces carbon emission while maintaining high accuracy. } First, we discuss \sol{}'s effectiveness over the baseline (Fig.~\ref{fig:eval1_clover}), and then, compare against other competing schemes in terms of accuracy and carbon emission savings (Fig.~\ref{fig:eval2_comnpare}).

First, from Fig.~\ref{fig:eval1_clover}, we observe that \sol{} yields over 75\% carbon emission savings across all applications with minimal accuracy degradation (2-4\%). When all three applications are hosted, \sol{} saves 80\% of carbon emission while keeping accuracy degradation at 3\% overall. This result is obtained via evaluation over a 48-hour span of varying carbon intensity using US CISO (California) trace. We note that different model variants can often degrade accuracy by over 10\% (Sec.~\ref{sec:motiv}). Therefore, na\"{\i}vely mixing different quality of models can lead to severe overall accuracy degradation with suboptimal carbon emission savings -- \sol{} avoids that situation by intelligently mixing models of different accuracy levels, carefully partitioning the GPUs, and adapting to the varying carbon intensity of the energy source. In fact, later in our evaluation, we demonstrate that one can configure \sol{} to set the accuracy loss to be as minimal as 0.2\%, and still achieve over 55\% carbon emission savings. 

Second, we observe that the service tail latency is also reduced, even though \sol{} does not explicitly optimize for this metric and only treats this as a constraint. This is because the use of lower-quality models with less computation allows us to reduce service latency. But, more interestingly, it also affirms \sol{} is effective at using lower-quality model variants to reduce service time despite relatively reduced GPU resources because of GPU partitioning. 

Next, Fig.~\ref{fig:eval2_comnpare} shows the comparison between \sol{} and other competing strategies. For each application, we show the accuracy gain and total carbon saved with respect to \base{}. The horizontal dashed line represents the best accuracy possible, while the vertical line represents the most carbon saving we can achieve. The intersection of these two extremes (the yellow star) represents the ambitious goal to achieve the best of both worlds. 

Fig.~\ref{fig:eval2_comnpare} confirms that \sol{} outperforms the competing schemes and is always closest to \oracle{}. As expected, the \opt{} scheme yields the highest carbon savings, but the lowest accuracy because it aggressively leverages the model variants and GPU partition to reduce carbon. \sol{} consistently outperforms \opt{} in terms of accuracy, while \sol{} is within 5\% of optimal carbon savings. We also note that the worst-case accuracy is represented by \opt{} case.

For the detection and language applications, \sol{} has outperformed \blover{} in terms of both accuracy and carbon. For the image classification task, \sol{} has similar accuracy as both \blover{} and \oracle{}, but its carbon saving is significantly higher than \blover{} and only slightly lower than \oracle{}. Note that the difference in the solution quality between \sol{} and \oracle{} is because \sol{}'s online search process declares maturity before finding the most optimal configuration that \oracle{}'s exhaustive offline search can find. 

We note that we chose to host a high-quality model on a GPU as a baseline because it aligns with current industry practices and keeps the average utilization still high. Nevertheless, even using different configurations as the baseline (e.g., second configuration instead of the first configuration in Fig.~\ref{fig:bkgd1_mig}), the key ideas of \sol{} are still beneficial and provide improvements.

Finally, we attempt to provide physical meaning and significance of \sol{}'s carbon savings (over 75\% compared to the baseline), using an example (e.g., 25 million inferences per day for AI services within US, which is still a conservative estimate as per~\cite{wu2022sustainable,patterson2022carbon}). To estimate the savings, we conservatively plug in our carbon saving number which is $6.77\times 10^{-3}$ g\coo{}/request. The US has an average carbon intensity of 380 g\coo{}/kWh~\cite{us-ci}, and we assume a datacenter has a PUE of 1.5. Using the back of the envelope estimations, we calculate that \sol{} can help save about 170 kg of \coo{} per day. This translates to the amount of carbon emitted by a gasoline car traveling 680 km or the amount of carbon saved by not burning 85 kg of coal every day~\cite{epa}. We acknowledge that these savings are only projections to gain a physical significance and estimate of potential savings by \sol{}-like solutions. We do not claim that they are directly related to a particular AI service (e.g., large language model inferences).

Next, we explain why \sol{} is able to outperform its competitors and operate close to \oracle{}. 

\begin{figure}
    \centering
    \includegraphics[scale=0.5]{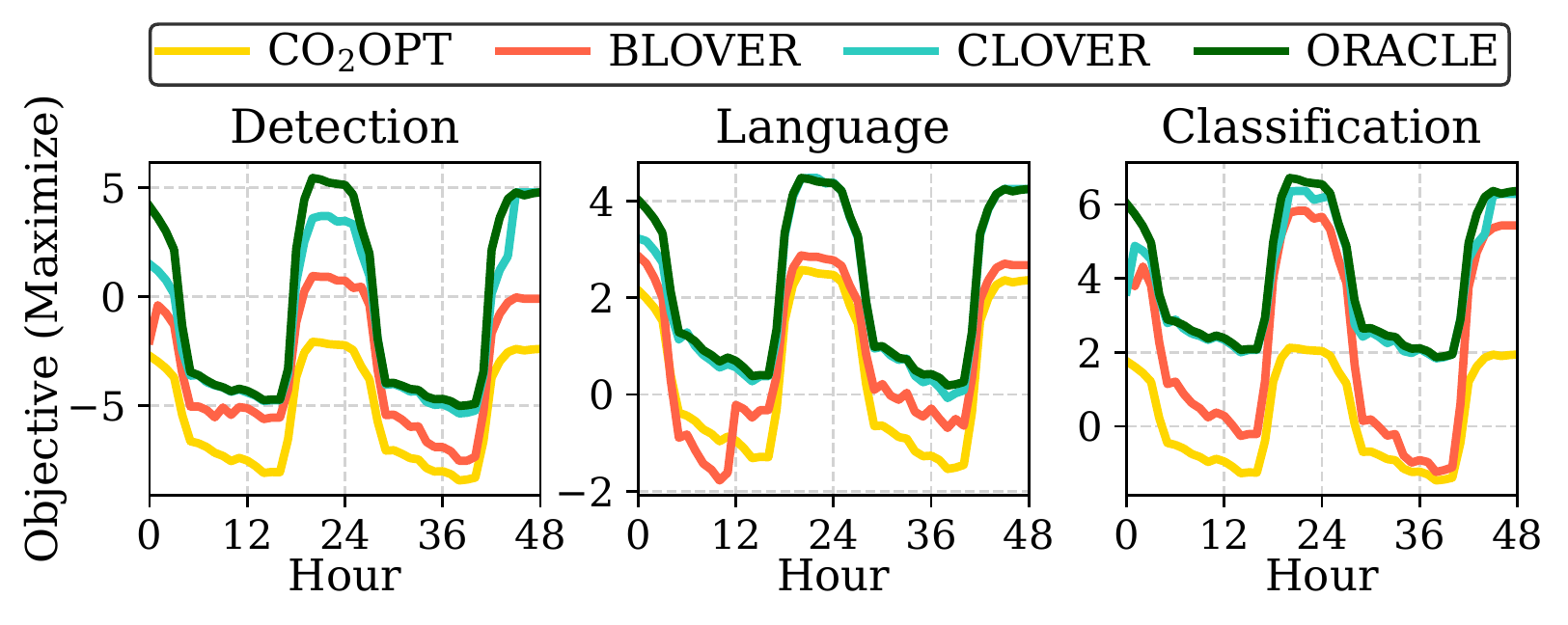}    
    \hrule
    \vspace{-0.4cm}    
    \caption{Over time, \sol{}'s optimization process automatically adjusts itself to maximize its optimization metric depending upon the carbon intensity. The outcome of \sol{}'s optimization process closely follows \oracle{} -- contributing to its high effectiveness.}
    \vspace{-0.5cm}
    \label{fig:eval3_score}
\end{figure}

\vspace{1.5mm}

\subsubsection{Why is \sol{} effective?}
\label{sec:eval_why}

\noindent\textbf{Sources of \sol{}'s effectiveness.} The achieved accuracy and carbon emission reduction are a result of  \sol{}'s objective function optimization (described in Section~\ref{sec:desi_formulate}). In order to gain a better understanding of the underlying factors, Fig.~\ref{fig:eval3_score} illustrates how various schemes optimize the objective function over time. It also visually shows the worst-case difference between \sol{} and \oracle{} (hour 0).

We make three major observations. Firstly, the outcome of \sol{}'s optimization process closely follows what the \oracle{} would have achieved, \sol{} overlaps with \oracle{}{} for the majority of the time. Secondly, the advantage of \sol{} is clear over \blover{} when visualizing the optimization objective. Note that \blover{} has the same termination condition as \sol{} but ends with worse configurations. This is because, without \sol{}'s graph-based optimization, \blover{} cannot quickly find a near-optimal configuration to keep up with the pace of the changing carbon intensity. But \blover{} is still better than \opt{} because, at least, it attempts to adapt to the carbon intensity. Lastly, we can compare \sol{}'s objective to \oracle{}'s at hours 0, 24, and 48. These timestamps have similar carbon intensities (Fig.~\ref{fig:eval0_ci}) and we can confirm that \sol{} gets more intelligent over time as it is getting closer and closer to \oracle{}. 

\vspace{2mm}

Next, we use the image classification application as an example to provide more detailed breakdowns of \sol{}'s effectiveness.

\begin{figure}
    \centering
    \includegraphics[scale=0.48]{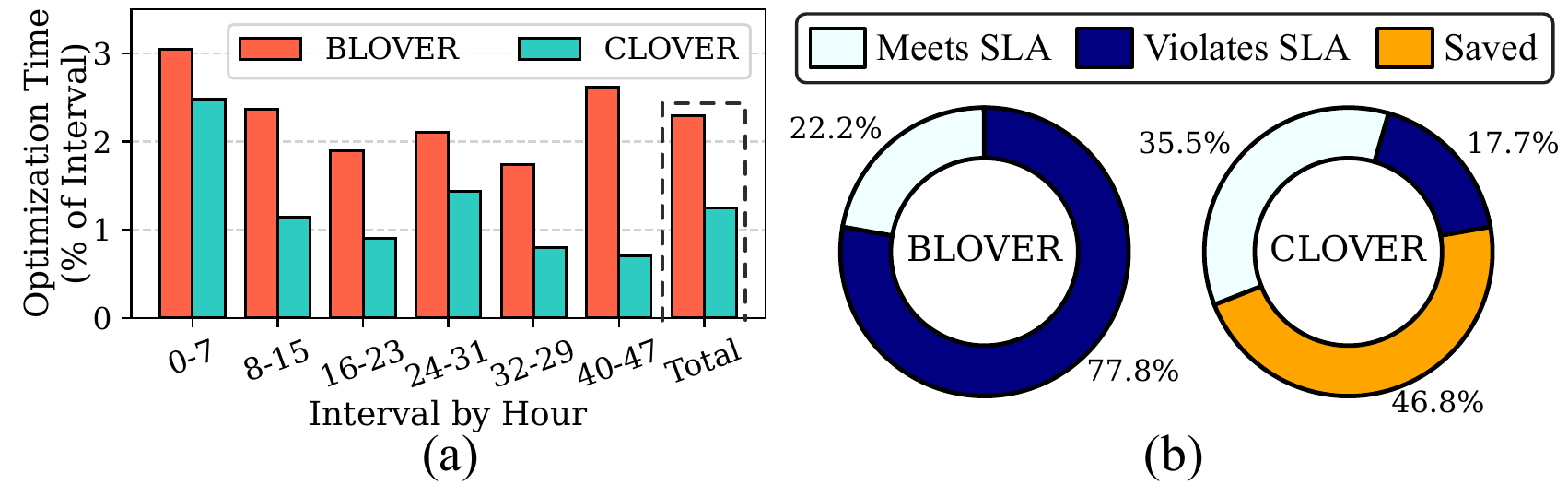} 
    \hrule
    \vspace{-0.3cm}    
    \caption{\sol{} spends only about 1.2\% of the total evaluated trace length (48 hours) in optimization, searching for near-optimal configurations (i.e., the optimization overhead is about 1.2\%). During the exploration, \sol{} explores less number of configurations (less than 50\% of \blover{}) and 60\% of all configurations evaluated by \sol{} meet the SLA constraint.}
    \label{fig:eval5_overhead}
\end{figure}

\sol{} performs optimization online that matches the quality of \oracle{}'s expensive offline solution. The online exploration process is present in both \blover{} and \sol{} and incurs some exploration overhead (this overhead is included in all the results of Sec.~\ref{sec:eval_eval}). In Fig.~\ref{fig:eval5_overhead} (a), we visualize the optimization time as a percentage of the total time span and show that \sol{}'s configuration-graph-based optimization saves a significant amount of time. The bars within the dashed box indicate that \blover{} spends 2.3\% of the total time running optimization, while \sol{} spends only 1.2\% of the time in optimization. By splitting the time span into six 8-hour windows, we can see that initially, both \blover{} and \sol{} spend more than 2.5\% of the time in optimization, but by the end, \sol{} becomes significantly more efficient than \blover{}. This improvement is due to the fact that \sol{} optimizes in a smaller graph space $x^g$ than the $(\bm{x}^p, \bm{x}^v)$ space (Sec.~\ref{sec:desi_graph}) .

In Fig.~\ref{fig:eval5_overhead} (b), we see that \sol{} outperforms \blover{} in avoiding SLA-violating configurations during optimization, resulting in a better user experience even during the optimization and reconfiguration phase. This is because \sol{} performs fewer evaluations (shown as "Saved") and the SA algorithm is able to guide \sol{} towards SLA-compliant graph neighborhoods.


\begin{figure}
    \centering
    \includegraphics[scale=0.5]{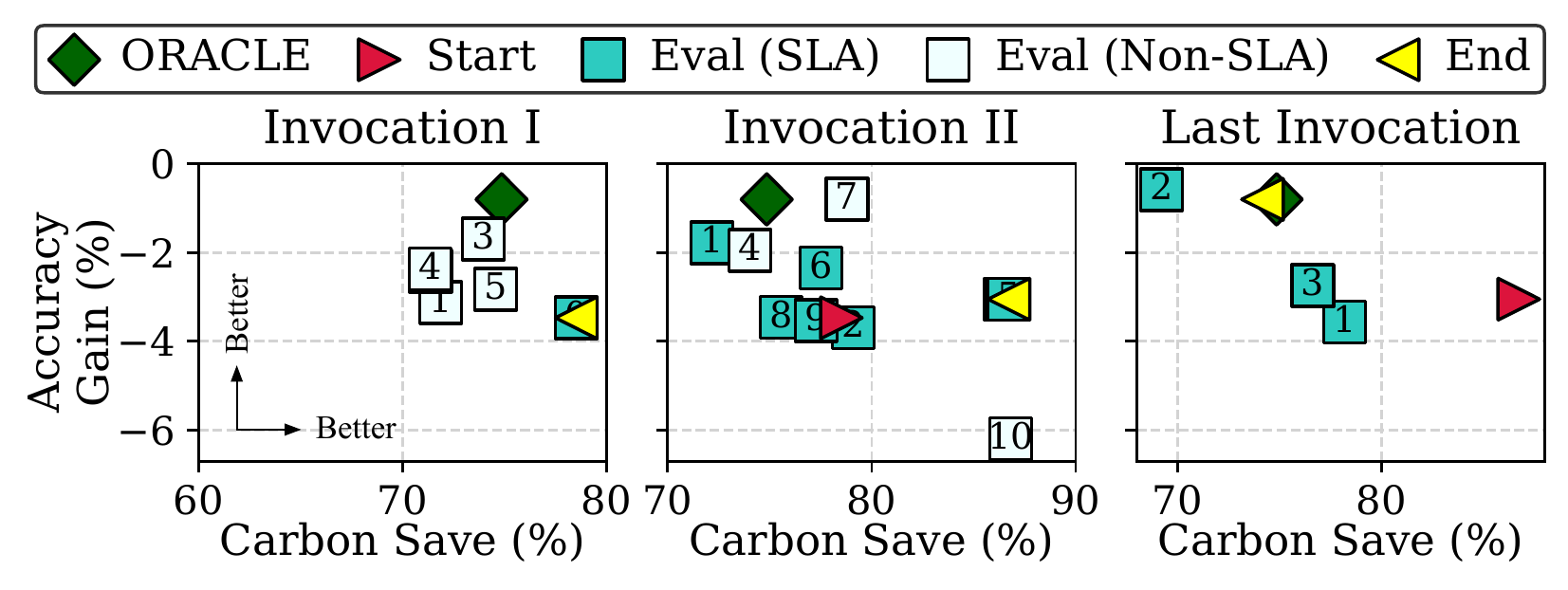} 
    \hrule
    \vspace{-0.4cm}    
    \caption{\sol{}'s optimization process successfully explores multiple promising configurations to find a balance between accuracy and carbon saving. \sol{} becomes significantly more intelligent compared to when it invoked the optimization for the first time.}
    \label{fig:eval6_transient}
\end{figure}

We configure \sol{} to invoke a new optimization process whenever \sol{} detects more than a 5\% change in the carbon intensity compared to the previous optimization run. For a visual illustration of further evidence, Fig.~\ref{fig:eval6_transient} shows the configurations that \sol{} evaluates for three sample invocation points -- the first, second, and last invocation in the trace.

In Fig.~\ref{fig:eval6_transient}, the configurations are labeled with numbers indicating the order of evaluation. For the first time \sol{} is deployed and invoked, it starts blindly thus most of the evaluated configurations cannot meet the SLA, and eventually, it has to settle with the only SLA-compliant configuration it has found (yellow triangle). When invocation (II) starts, its initial configuration (red triangle) is invocation (I)'s best configuration found. This optimization run is much more effective because most of its configurations are SLA-compliant, and has settled on a configuration (yellow triangle) that has both higher accuracy and carbon savings than its initial configuration. Finally, in the last invocation of our trace span, \sol{} is able to converge to \oracle{} with only four evaluations, all configurations being SLA-compliant. 






\subsubsection{\sol{}'s robustness and adaptivity.}
\label{sec:eval_adapt}

\begin{figure}
    \centering
    \includegraphics[scale=0.48]{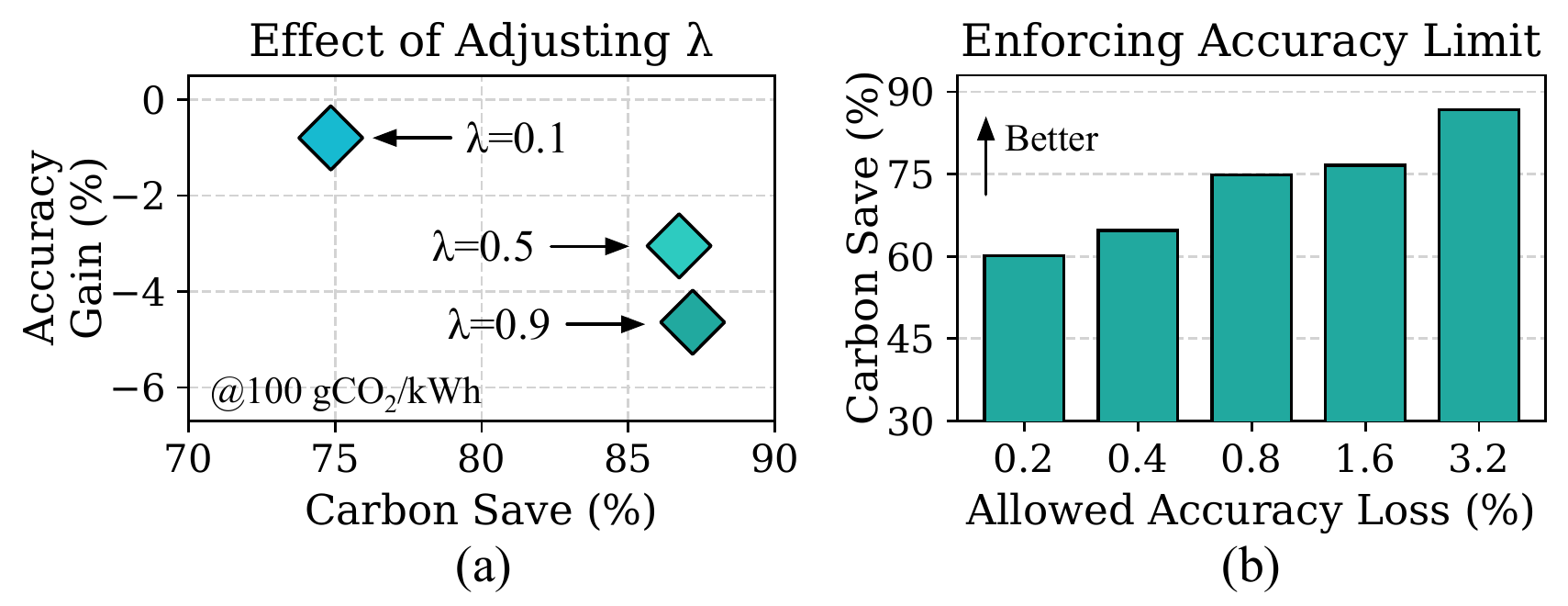}    
    \hrule
    \vspace{-0.4cm}    
    \caption{\sol{} can achieve different levels of accuracy and carbon savings by adjusting the weight factor on different objectives (lower $\lambda$ parameter yields higher accuracy). \sol{} excels at achieving high carbon savings when accuracy constraints are enforced.} 
    \label{fig:eval7_acc}
\end{figure}

\vspace{3mm}

\noindent\textbf{\sol{} is effective under different scenarios and constraints, including accuracy thresholds and geographies/carbon intensity.} Recall that \sol{}'s $\lambda$ parameter in Eq.~\ref{eq:desi_obj} allows customizing a trade-off between accuracy and carbon savings. Fig.~\ref{fig:eval7_acc} demonstrates the effectiveness of \sol{} when $\lambda$ shifts from 0.1 to 0.9 (at 100 g\coo{}/kWh intensity). As expected, \sol{} gradually trades more accuracy for higher carbon savings. Next, we show that the providers can specify accuracy loss as a constraint to \sol{}. In such cases, \sol{} guarantees that a maximum threshold of accuracy is allowed to participate in the trade-off. Fig.~\ref{fig:eval7_acc} (b) shows that even with a 0.2\% to 0.8\% accuracy loss compared to \base{}, \sol{} can still provide 60\% to 75\% total carbon saving.

\begin{figure}
    \centering
    \includegraphics[scale=0.50]{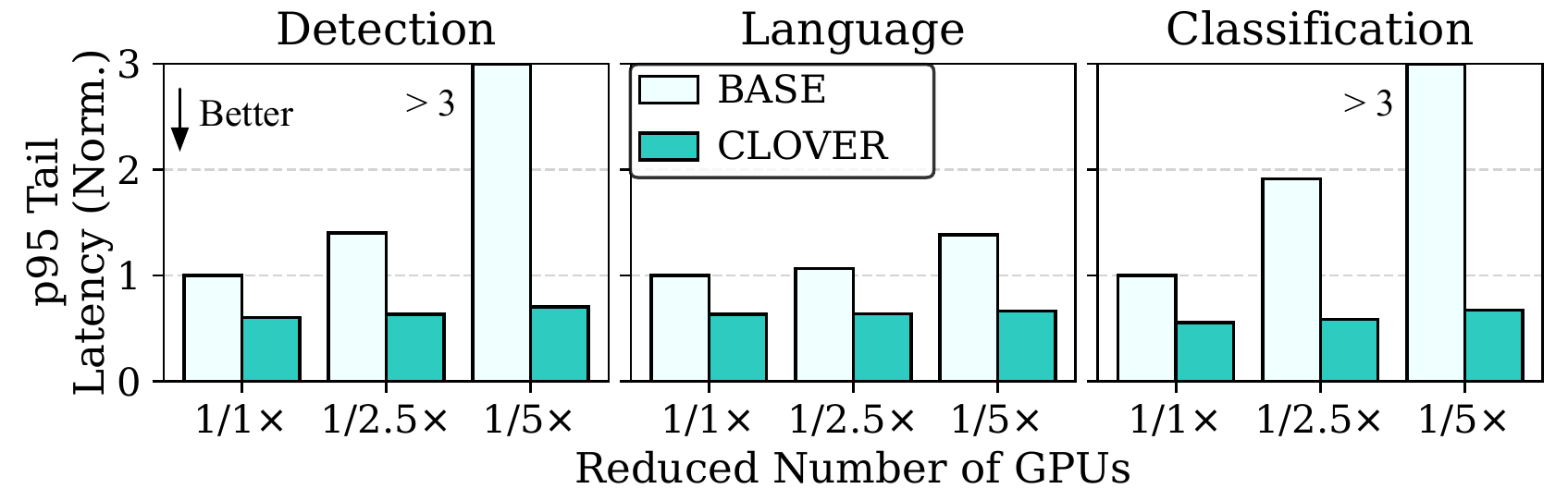}    
    \hrule
    \vspace{-0.4cm}    
    \caption{\sol{}'s co-location and mixed-quality serving enable reductions in the number of GPUs while meeting SLA targets.} 
    \vspace{-0.5cm}
    \label{fig:eval9_shrink}
\end{figure}

\vspace{3mm}
Next, we show in Fig.~\ref{fig:eval9_shrink} that \sol{} enables us to provision a lesser number of GPUs to meet the same SLA target as \base{} while saving carbon, this is because of intelligent GPU resource partitioning and resource sharing among mixed-quality models. In Fig.~\ref{fig:eval9_shrink}, the x-axis labels represent provisioning 10, 4, and 2 GPUs respectively, and we compare the service tail latency normalized to the non-partitioned baseline with the highest quality models and ten GPUs. The violated SLA for \base{} scheme ($>1$) implies provisioning 10 GPUs is necessary for \base{}, but when \sol{} is employed, even 2 GPUs can be sufficient to meet the same service goals because \sol{} is more efficient in utilizing the GPU hardware and leverages mixed-quality models. The takeaway is that as \sol{} explicitly reduces the operational carbon emission, it can also implicitly reduce the carbon emission incurred in manufacturing, transporting, and cooling of the unneeded server machines. 

\begin{figure}
    \centering
    \includegraphics[scale=0.50]{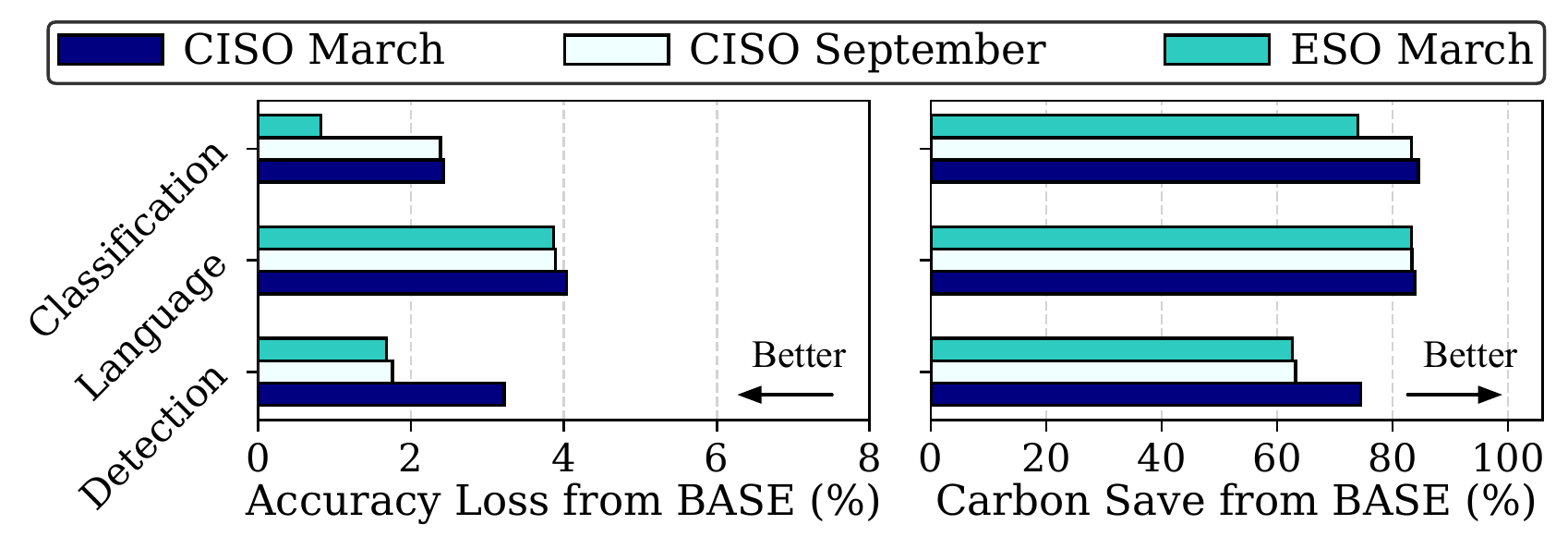}    
    \hrule
    \vspace{-0.4cm}    
    \caption{\sol{} is effective across different geographical regions and seasons with varying carbon intensity.} 
    \label{fig:eval8_sense}
\end{figure}

Finally, Fig.~\ref{fig:eval8_sense} confirms that \sol{} remains effective when running at different geographical locations and during different seasons -- offering over 60\% carbon savings with limited accuracy loss across all applications.

\section{Related Work}
\label{sec:relat}

\noindent\textbf{Rising interests in carbon-friendly systems.} Totally Green~\cite{chang2012totally} was among the first studies to provide system-focused modeling of environmental impact during datacenter production and operation. Recently, there has been a rise in interest in addressing the carbon emission of large-scale systems as IT companies aim to achieve net zero goals. ACT~\cite{gupta2022act} and Chasing Carbon~\cite{gupta2022chasing} provide in-depth industry-reported \coo{} characterizations to build carbon modeling tools for desktops and mobile phones. Although these works do not provide carbon-friendly inference solutions, our motivation to build \sol{} is driven by their industry-grade numbers. Carbon Explorer~\cite{acun2023carbon} and CICS~\cite{radovanovic2022carbon} make datacenters carbon-aware by greedily pausing or delaying batch jobs in a datacenter when there is a limited supply of renewable energy (potentially violating SLAs) -- \sol{} hosts services that must be available 24/7 and achieves carbon savings without compromising service capacity and SLA. 

Sustainable AI~\cite{wu2022sustainable} presents an end-to-end analysis of how Meta uses hardware-software co-design to reduce its AI carbon footprint. Similar efforts toward reducing carbon emission also appear in Google and Microsoft~\cite{patterson2021carbon,patterson2022carbon,anderson2022treehouse,dodge2022measuring} related to ML training and software development, but none of these prior works address the challenge of making ML inferences more carbon-friendly via GPU partitioning and mixed-quality models. One closely relevant  work~\cite{dodge2022measuring} has attempted simple solutions of pausing cloud instances at high carbon intensity – which is not possible for SLA-critical ML inference. \sol{} is motivated by these prior works, but takes a significant step forward – it designs and builds the first real-system framework to reduce the carbon footprint of ML inference services.

\vspace{3mm}
\noindent\textbf{Machine learning inference service.} Many previous research works have focused on various aspects of ML inference services including latency predictability~\cite{258862,cui2021enable,xu2022igniter}, cloud cost efficiency~\cite{romero2021llama,li2021ribbon,hu2021scrooge,liu2021jizhi,li2023kairos}, and adaptive query batching~\cite{fu2022qos,ali2020batch,choi2021lazy,crankshaw2017clipper}.  \sol{} distinguishes itself from these previous contributions, as it focuses on building a carbon-aware inference system.

Some works have focused on the energy perspective of ML inference on GPU clusters~\cite{kang2022cost,komoda2013power,nabavinejad2021batchsizer} and battery-powered devices~\cite{xu2020approximate,kannan2021budget,zadeh2020gobo}. In particular, prior works~\cite{kang2022cost,komoda2013power,nabavinejad2021batchsizer} are useful in terms of making DNN inference more energy efficient. But, they do not consider the carbon intensity, the accuracy-carbon trade-off, and automatic carbon-aware model variant and GPU partition configuration, which are the key insights of \sol{}. These prior works do not design and build a framework to reduce the carbon footprint of SLA-constrained ML inference services.

\vspace{3mm}

INFaaS~\cite{romero2021infaas} builds a model-less service that hides the model selection and hardware from developers but does not focus on carbon savings or hardware sharing for dynamically adapting to a changing environment. Gslice~\cite{dhakal2020gslice} and Gpulet~\cite{choi2022serving} improve GPU utilization using NVIDIA Multi-Process Service (MPS)~\cite{nvidia-mps}, but MPS does not provide performance isolation, memory protection, and error isolation while \sol{} uses NVIDIA's latest MIG partition that provides hardware-based isolation and protections. MISO~\cite{li2022miso} uses MIG partition to improve system throughput, but it does not consider accuracy/carbon/SLA of inference services while \sol{} designs a novel configuration-graph-based optimization to optimize the MIG partition and service model online. Among all the works in this field, \sol{} is the first inference framework that leverages mixed model variants and MIG GPU partition to become carbon-aware.



\section{Conclusion}
\label{sec:conclusion}
This paper has presented a novel approach, \sol{}, to address the AI sustainability challenge by designing a carbon-aware ML inference service framework. \sol{} provides an intelligent and practical solution to balance the trade-offs between carbon emission, accuracy, and SLA targets. We believe that our work will encourage researchers and practitioners to develop richer GPU partitioning hardware support and generate mixed-quality models to reduce the inference carbon footprint further. We hope that \sol{} will inspire the community to work toward more environmentally sustainable solutions for ML inference systems.



\begin{acks}
This material is based upon work supported by the Assistant Secretary of Defense for Research and Engineering under Air Force Contract No. FA8702-15-D-0001, and United States Air Force Research Laboratory Cooperative Agreement Number FA8750-19-2-1000. Any opinions, findings, conclusions, or recommendations expressed in this material are those of the author(s) and do not necessarily reflect the views of the Assistant Secretary of Defense for Research and Engineering, or the United States Air Force. The U.S. Government is authorized to reproduce and distribute reprints for Government purposes notwithstanding any copyright notation herein. 
\end{acks}

\bibliographystyle{unsrtnat}
\bibliography{refs}

\end{document}